\documentclass[a4paper,11pt]{article}
\pdfoutput=1

\usepackage{amssymb,mathrsfs,amsmath}
\usepackage{a4wide}
\usepackage{color,xcolor}
\usepackage{slashed,soul}
\usepackage{graphicx}
\usepackage{amsfonts}
\usepackage{lscape}
\def\linkcolor{cyan!70!black}

\usepackage[
colorlinks=true
,urlcolor=\linkcolor
,anchorcolor=\linkcolor
,citecolor=\linkcolor
,filecolor=\linkcolor
,linkcolor=\linkcolor
,menucolor=\linkcolor
,linktocpage=true
,pdfproducer=medialab
,pdfa=true
]{hyperref}
\usepackage{amsthm}
\usepackage{booktabs}
\usepackage{array}
\usepackage{rotating}
\usepackage[numbers, sort&compress]{natbib}
\usepackage{multirow}
\usepackage{float}
\usepackage[utf8]{inputenc}
\usepackage[T1]{fontenc}
\usepackage{appendix}
\usepackage{epsfig}

\newcommand{\be}{\begin{equation}}
\newcommand{\ee}{\end{equation}}

\newcommand{\beq}{\begin{equation}} 
\newcommand{\eeq}{\end{equation}} 
\newcommand{\ba}{\begin{array}}  
\newcommand{\ea}{\end{array}} 
\newcommand{\bea}{\begin{eqnarray}}  
\newcommand{\eea}{\end{eqnarray} }  
\newcommand{\bal}{\begin{align}}
\newcommand{\eal}{\end{align}}   
\newcommand{\bi}{\begin{itemize}}  
\newcommand{\ei}{\end{itemize}}  
\newcommand{\ben}{\begin{enumerate}}  
\newcommand{\een}{\end{enumerate}}  
\newcommand{\bc}{\begin{center}}
\newcommand{\ec}{\end{center}} 
\newcommand{\bt}{\begin{table}}
\newcommand{\et}{\end{table}}  
\newcommand{\btb}{\begin{tabular}}
\newcommand{\etb}{\end{tabular}}

\newcommand{\Tr}[1]{\text{Tr}\left\{#1\right\}}
\newcommand{\abs}[1]{\left\lvert#1\right\rvert}

\newcommand{\imag}[1]{\text{Im}\left[ #1\right]}
\newcommand{\mdiag}{m^{\text{diag}}}
\newcommand{\EWvev}{v}
\newcommand{\mutilde}{\tilde{\mu}}
\newcommand{\mlightest}{m_{\text{lightest}}}
\newcommand{\UPMNS}{U_{\text{PMNS}}}
\newcommand{\deltaCP}{\delta_{\text{CP}}}

\allowdisplaybreaks

\let\OLDthebibliography\thebibliography
\renewcommand\thebibliography[1]{
  \OLDthebibliography{#1}
  \setlength{\parskip}{0pt}
  \setlength{\itemsep}{0pt plus 0.3ex}
}

\begin{document}


\begin{titlepage}

\begin{flushright}
IFT-UAM/CSIC-22-104
 \end{flushright}
\vspace{0.2truecm}

\begin{center}
\renewcommand{\baselinestretch}{1.8}\normalsize
\boldmath
{\LARGE\textbf{
HNL mass degeneracy: implications for \\ low-scale seesaws, LNV at colliders and leptogenesis
}}
\unboldmath
\end{center}

\vspace{0.4truecm}

\renewcommand*{\thefootnote}{\fnsymbol{footnote}}

\begin{center}

{\bf 
Enrique Fern\'andez-Mart\'inez\footnote{\href{mailto:enrique.fernandez-martinez@uam.es}{enrique.fernandez-martinez@uam.es}}, 
Xabier Marcano\footnote{\href{mailto:xabier.marcano@uam.es}{xabier.marcano@uam.es}}
and Daniel Naredo-Tuero\footnote{\href{mailto:daniel.naredo@uam.es}{daniel.naredo@uam.es}}
}

\vspace{0.5truecm}

{\footnotesize

Departamento de F\'{\i}sica Te\'orica and Instituto de F\'{\i}sica Te\'orica UAM/CSIC,\\
Universidad Aut\'onoma de Madrid, Cantoblanco, 28049 Madrid, Spain
}

\vspace*{2mm}
\end{center}

\renewcommand*{\thefootnote}{\arabic{footnote}}
\setcounter{footnote}{0}

\begin{abstract}
\noindent 
Low-scale seesaw variants protected by lepton number symmetry provide a natural explanation of the smallness of neutrino masses but, unlike their higher-scale counterparts, with potentially testable phenomenology. The approximate lepton number symmetry arranges the heavy neutrinos in pseudo-Dirac pairs, which might be accessible at collider or even beam dump experiments if their mass is low enough and their mixing with the active neutrinos sufficiently large. Despite their pseudo-Dirac nature, their small mass splittings may lead to oscillations that prevent the cancellation of their potential lepton-number-violating signals. Interestingly, these small splittings may also resonantly enhance the production of a lepton number asymmetry for low-scale leptogenesis scenarios or, for extremely degenerate states, lead to an asymmetry large enough to resonantly produce a keV sterile neutrino dark matter candidate with the correct relic abundance via the Shi-Fuller mechanism. 
In this work we explore the parameter space of the different low-scale seesaw mechanisms and study the size of these splittings, given their important and interesting phenomenological consequences.
While all low-scale seesaw variants share the same dimension 5 and 6 operators when integrating out the heavy states, we point out that the mass splitting of the pseudo-Dirac pairs are very different in different realizations such as the inverse or linear seesaw. 
This different phenomenology could offer a way to discriminate between low-scale seesaw realizations.  
 
\end{abstract}

\end{titlepage}

\tableofcontents

\section{Introduction}

Among the different extensions of the Standard Model (SM) of particle physics able to accommodate the evidence for neutrino masses and mixings from the neutrino oscillation phenomenon, the inclusion of right-handed neutrinos is arguably the simplest. Given their singlet nature, a Majorana mass at a new energy scale at which lepton number ($L$) is violated is allowed for them together with standard Yukawa couplings with the SM left-handed neutrinos. 

In the high-scale seesaw mechanism~\cite{Minkowski:1977sc,Mohapatra:1979ia,Yanagida:1979as,Gell-Mann:1979vob}, the new mass scale is much larger than the electroweak scale, suppressing the masses of the mostly active neutrino mass eigenstates and providing a rationale for the smallness of neutrino masses. Nevertheless, that same explanation of the smallness of neutrino masses renders the mechanism virtually untestable, since also the mixing of the new states with the active neutrinos is similarly suppressed and their energy scale too high to probe. Moreover, such a high new-physics scale coupled to the Higgs through the neutrino Yukawa couplings introduces a hierarchy problem~\cite{Vissani:1997ys,Casas:2004gh}.

Alternatively, the smallness of neutrino masses may also be naturally explained through a symmetry argument. Indeed, the Weinberg $d=5$ operator~\cite{Weinberg:1979sa} obtained upon integrating out the new heavy states and that gives rise to the masses of the mostly active neutrinos violates $L$ by two units, being a Majorana type of mass. Thus, an approximate $L$ symmetry imposed in the Lagrangian would naturally suppress at all orders the $d=5$ operator and hence neutrino masses~\cite{Branco:1988ex,Kersten:2007vk, Abada:2007ux}. On the other hand, the only $d=6$ operator obtained at tree level upon integrating out the heavy states and that encodes the heavy-active neutrino mixing through deviations of the lepton mixing matrix from Unitarity~\cite{Broncano:2002rw}, does not violate $L$. Hence, in symmetry-protected versions of the seesaw mechanism, the $d=6$ operator is actually expected to be less suppressed than its $d=5$ counterpart and the heavy-active mixing can be sizable with no conflict with the small neutrino masses. Given the approximate $L$ symmetry the heavy neutrinos arrange in pseudo-Dirac pairs. The symmetry suppression allows these pairs to be relatively light and have sizable mixings, allowing to probe for the existence of these heavy neutral leptons (HNLs) at collider or beam dump experiments~\cite{Abdullahi:2022jlv}, while the masses of the mostly active neutrinos from the symmetry-protected Weinberg operator remain small.

Several variants of these low-scale, symmetry-protected seesaws have been studied depending on the source of the small symmetry-breaking terms, such as the inverse~\cite{Mohapatra:1986aw,Mohapatra:1986bd} and linear~\cite{Akhmedov:1995ip,Malinsky:2005bi} seesaws. All of them are characterised by small neutrino masses from the $d=5$ Weinberg operator proportional to the small, symmetry-breaking parameters and by large mixing of the heavy pseudo-Dirac pairs with the active neutrinos encoded in the $d=6$ operator. However, the different possible sources of $L$ violation that give rise to the light neutrino masses do lead to significantly different splittings among the masses of the members of the pseudo-Dirac pairs~\cite{Antusch:2017ebe,Drewes:2019byd}. Indeed, even though both the Weinberg operator and the heavy neutrino mass splitting violate $L$ by two units, the relation between these two quantities is different in the different realizations of low-scale seesaws.

Interestingly, the mass splitting of the pseudo-Dirac pairs has very significant phenomenological consequences. On the one hand, depending on its size, these splittings may drive oscillations between two states provided that the oscillation length is smaller than the decay length. In such a scenario, the cancellation between the opposing Majorana phases of the two members of the pair, which strongly suppresses any $L$-violating process given their pseudo-Dirac nature, is prevented~\cite{Drewes:2019byd,Gluza:2015goa,BhupalDev:2015kgw,Anamiati:2016uxp,Das:2017hmg,Antusch:2017ebe}. In this regime, the decay products of the heavy neutrinos would instead correspond to those of a Majorana state. The areas of the parameter space for which the heavy neutrinos behave as Dirac or Majorana thus strongly depend on the mass splitting between the two members of the pseudo-Dirac pairs.

On the other hand, when resonant~\cite{Pilaftsis:2003gt} or ARS leptogenesis~\cite{Akhmedov:1998qx,Asaka:2005pn} is considered in low-scale seesaws, a strong degeneracy among the sterile neutrino states can lead to a significant enhancement of the $L$ asymmetry produced (see also~\cite{Canetti:2010aw,Garny:2011hg,Canetti:2012kh,Abada:2015rta,Hernandez:2015wna,Hernandez:2016kel,Drewes:2016jae,Abada:2017ieq,Drewes:2017zyw,Eijima:2018qke,Abada:2018oly,Klaric:2021cpi,Drewes:2022kap,Hernandez:2022ivz} and references therein).

The degree of degeneracy is thus fundamental to asses the size of the final baryon asymmetry. Furthermore, while the bounds on the mixing of a kev-scale, sterile neutrino dark matter candidate from X ray constraints are too strong to allow its production via mixing through the Dodelson-Widrow mechanism~\cite{Dodelson:1993je}, it has been shown that this production can be resonantly enhanced if in the background of a very large $L$ asymmetry through the so-called Shi-Fuller mechanism~\cite{Shi:1998km,Ghiglieri:2019kbw,Ghiglieri:2020ulj}. The generation of such a large background $L$ asymmetry in turn requires an even stronger degeneracy among the pseudo-Dirac pairs. Interestingly, a very large $L$ asymmetry would also help to reconcile recent estimations of the primordial deuterium~\cite{Workman:2022ynf} and $^4$He~\cite{Matsumoto:2022tlr} abundances which show a $\sim 2 \sigma$ and $\sim 3 \sigma$ tension respectively with the predictions of Big Bang Nucleosynthesis (BBN) in the standard $\Lambda$CDM model~\cite{Burns:2022hkq}.

Therefore, the level of degeneracy of the heavy neutrinos is critical, not only for the potential observation of $L$ violating signals if the heavy neutrinos are eventually produced at experimental facilities, but also for the potential generation of the baryon asymmetry of the universe and of sterile neutrino dark matter. 
In this work we perform a thorough scan of the full parameter space of the different realizations of the low-scale seesaws, studying the allowed size of the mass splitting between members of the pseudo-Dirac pairs. We start by reviewing in section~\ref{Sec:2HNL} the original results of Refs.~\cite{Antusch:2017ebe,Drewes:2019byd}, which focus on the minimal scenario with only one pair of HNLs, highlighting the value of the HNL mass splittings and its connection with light neutrino masses. In this restricted scenario there is a one to one correspondence between the HNL and light neutrino mass splittings. Our main original contribution in this study is to go beyond these minimal setups and explore how much the naive expectations for the value of the HNL mass splitting in the minimal setup can change when more than a single HNL pair is considered. This thorough study is performed for a linear seesaw in section~\ref{Sec:LSS} and an inverse seesaw in section~\ref{Sec:ISS}. In both cases, we generalise the naive estimations obtained in the minimal scenario, showing and quantifying under which conditions is it possible to find largely enhanced o suppressed HNL mass splittings. Thus we conclude that the possible values of the HNL splitting can be significantly beyond those studied so far, but only if certain cancellations or symmetries that we carefully quantify and comment upon take place. 
Our main original results are summarized in section~\ref{Sec:Discussion} through the comparison of Figures~\ref{FeelingPlot} and~\ref{BoundaryLNV_2HNL}, which depict respectively the expectation for the HNL mass splitting and LNV phenomenology in minimal low-scale seesaws from Refs.~\cite{Antusch:2017ebe,Drewes:2019byd} and how that expectation can change when more than a single pseudo-Dirac pair is present, as well as the conditions required to reach the different values. These results also show how separated the predictions between the linear and inverse realizations can be, so that it might be used as a way of discerning between the mechanisms for neutrino masses and mixings.



\section{The minimal 2-HNL scenario and its phenomenology}
\label{Sec:2HNL}
The observation of neutrino flavour change and its explanation through the oscillation phenomenon requires two distinct mass splittings among the three neutrino mass eigenstates. These are the so-called ``atmospheric'' and ``solar'' mass splittings, which correspond to $|\Delta m^2_{31}| = 2.5 \cdot 10^{-3}$ eV$^2$ and $\Delta m^2_{21} = 7.4 \cdot 10^{-5}$ eV$^2$, respectively~\cite{Esteban:2020cvm}. Thus, at least two of the three light neutrino mass eigenstates need to acquire non-zero masses, which in turn requires the addition of at least two HNLs in the type-I seesaw. These heavy neutrinos $N$ will, in all generality, have two contributions to their masses. The usual Yukawa couplings with the SM left-handed leptons leading to a Dirac mass term $m_D$ after the Higgs develops its vacuum expectation value $\EWvev$; and a symmetric Majorana mass $M_M$, which is allowed for them at the Lagrangian level given their gauge singlet nature. Thus, the mass terms can be combined in a single mass matrix $\overline{\mathcal{N}} \mathcal{M}_\nu \mathcal{N}^c$ where:
\begin{equation}
    \mathcal{M}_\nu=\begin{pmatrix}
        0&m_D\\m_D^T&M_M
    \end{pmatrix}\,,
    \label{type1SSmassmatrix}
\end{equation}
and $\mathcal{N}=\left(\nu_L\hspace{0.1cm}N^c\right)^T $. In the seesaw limit of $m_D \ll M_M$, this matrix can be block-diagonalised through a unitary transformation $V$ 
\begin{equation}
    \mathcal{M}_\nu^{\text{block}}=V^T \mathcal{M}_\nu V=\begin{pmatrix}
         m_\nu&0\\0&M_N
     \end{pmatrix}\,,
     \label{type1SSBlockDiagonalisation}
\end{equation}
where
\begin{equation}
    V=\begin{pmatrix}
		1-\frac{1}{2}\Theta^*\Theta^T&\Theta^*\\
		-\Theta^T&1-\frac{1}{2}\Theta^T\Theta^*
	\end{pmatrix}+O(\Theta^3)\,,
	\label{type1SSVmatrix}
\end{equation}
and $\Theta=m_D M_M^{-1}$ represents the mixing between the active SM neutrinos and the HNLs. 
This rotation gives for $m_\nu$ the well-known coefficient for the Weinberg $d=5$ operator in the type-I seesaw:
\begin{equation}
    m_\nu\simeq-\Theta M_M \Theta^T\,,
    \label{type1SSmnu}
\end{equation}
and for the heavy neutrinos: 
\begin{equation}
    M_N=M_M+\dfrac{1}{2}\left(\Theta^\dagger\Theta M_M+M_M\Theta^T\Theta^*\right)\,.
    \label{type1SSMN}
\end{equation}
These two symmetric matrices can finally be diagonalised via Unitary transformations $U'$ and $V'$, respectively, so that the full rotation relating the flavour and mass eigenstates is given by:
\begin{equation}
U=\begin{pmatrix}
		 \left(1-\frac{1}{2}\Theta^*\Theta^T\right) U' & \Theta^* V' \\
		 - \Theta^T U'& \left(1-\frac{1}{2}\Theta^T\Theta^* \right) V'
	\end{pmatrix}+O(\Theta^3)\,.
	\label{Eq:fullmatrix}
\end{equation}
Notice that the upper-left $3\times3$ sub-block that will appear in the CC interactions of the light neutrinos with the charged leptons, that is the PMNS mixing matrix, is not unitary. The unitarity deviation $\Theta^*\Theta^T/2$ corresponds to the coefficient of the only $d=6$ operator obtained at tree level when integrating out the HNLs~\cite{Broncano:2002rw}. Conversely, the upper-right block of Eq.~\eqref{Eq:fullmatrix}, $U_{\ell N}$, contains the coupling of the heavy neutrinos with charged leptons via CC interactions, for which important experimental constraints from direct searches exist~\cite{Bolton:2019pcu,MatheusRepository}.

The minimal scenario with only 2 HNLs, focusing in particular on its low-scale realization where the HNLs form a pseudo-Dirac pair, was studied in depth in Ref.~\cite{Gavela:2009cd}. Here we will review its allowed parameter space when the correct pattern of masses and mixings of the light neutrinos is recovered and discuss in particular the expected mass splitting of the pseudo-Dirac pair and its correlation to the observed light neutrino mass splittings. 

In all generality, the mass matrix containing the possible Yukawa couplings between the SM neutrinos and the two HNLs as well as a Majorana mass for the HNLs can be written as
\begin{equation} \label{NeuMass}
    \mathcal{M}_\nu =
    \begin{pmatrix}
        0& Y\EWvev/\sqrt{2}&  Y'\EWvev/\sqrt{2}\\
         Y^T \EWvev/\sqrt{2}& \mu' &M\\
        Y'^T\EWvev/\sqrt{2} &M& \mu
    \end{pmatrix},
\end{equation}
and $\mathcal{N}=\left(\nu_L\hspace{0.1cm}N^c\hspace{0.1cm}N'\right)^T $, with $N$ and $N'$ the right- and left-handed components of the would-be pseudo-Dirac pair, respectively. In order for this general type-I seesaw to lead to small neutrino masses for the mostly SM neutrino mass eigenstates but at the same time allow for sizable Yukawas and a low-mass scale for the HNLs, an approximate $L$ symmetry must be imposed~\cite{Kersten:2007vk,Abada:2007ux,Moffat:2017feq}, so that $\mu/M, \mu'/M, Y' \ll 1$.  Indeed, in the limit of $\mu=\mu'=Y'=0$, $L$ is exactly conserved and the HNLs form a Dirac pair of mass $M$. However, the mixing of the SM neutrinos with the left-handed component of the Dirac HNL ($N'$), and thus its associated phenomenology, can still be sizable. In particular:
\begin{equation}
\Theta = \left(0  \hspace{0.2cm} \frac{Y \EWvev}{\sqrt{2} M} \right) \equiv \left(0  \hspace{0.2cm} \theta \right)\,,
\end{equation}
is unsuppressed by any small parameter as the $d=6$ operator is not protected by $L$. 

When introducing a $L$ violating term to generate neutrino masses, the same parameters will also induce a non-zero splitting $\Delta M$ between the two members of the pseudo-Dirac pair, pointing towards a connection between active neutrino masses and HNL mass splittings.
The explicit relation between the two depends however on the particular low-scale realization considered. 
The aim of this work is precisely to explore in detail this connection. As a first step, we devote this section to first sketch the naive expectations for different models.

The different variants of low-scale seesaws induce small neutrino masses by introducing some of the soft breaking terms in Eq.~\eqref{NeuMass}. In particular, the inverse seesaw (ISS) introduces a small Majorana mass $\mu\ll M$ for the $N'$ states, while the linear seesaw (LSS) considers a small Yukawa coupling $Y'$ between the SM lepton doublet and $N'$.
Their contribution to the light neutrino masses reads: 
\begin{equation}
m_\nu \simeq \dfrac{\EWvev^2 \mu}{2M^2} Y Y^T -\dfrac{\EWvev^2}{2M} \left(Y Y'^T+Y'Y^T\right).
\label{eq:2Nmass}
\end{equation}

Finally, we could also introduce a small Majorana mass $\mu'\ll M$ for the $N$ states, although it does not contribute to the light neutrino masses at tree level, since it does not increase the rank of $\mathcal M_\nu$.
Nevertheless, as it does break $L$, neutrino masses are no longer protected and it contributes at one loop\footnote{Notice that the logarithmic dependence in the loop contribution has been expanded assuming $M \gg M_H, M_Z, \mu'$ as in Ref.~\cite{Lopez-Pavon:2012yda}, but, for our all our numerical computations, only $M \gg \mu'$ will be assumed.}~\cite{Pilaftsis:1991ug,Lopez-Pavon:2012yda}:
\begin{equation}
m_\nu \simeq \dfrac{\mu'}{16 \pi^2} \dfrac{M_H^2 + 3 M_Z^2}{2 M^2} Y Y^T.
\label{eq:loopISS}
\end{equation}
This term has the same structure than the ISS contribution in Eq.~\eqref{eq:2Nmass}, so we will refer to this realization as {\it loop-induced ISS}.

Notice that in this case with only 2 HNLs, the ISS contributions, both at tree level through $\mu$ in Eq.~(\ref{eq:2Nmass}) and at loop level via $\mu'$ in Eq.~(\ref{eq:loopISS}), have rank 1 and hence can only give mass to a single light neutrino. 
Conversely, the LSS contribution with $Y'$ has rank 2 and may explain the observed pattern of neutrino masses and mixings. 
Moreover, the ISS contributions can be combined with the LSS by redefining an ``effective'' Yukawa coupling
$
\hat{Y}'= Y' - \left(\mu/2 M + \mu'\left(M_H^2 + 3 M_Z^2\right)/32 \pi^2 M \right)Y\,,
$
and their contributions to $m_\nu$ can be all combined as
\begin{equation}
m_\nu = -\dfrac{\EWvev^2}{2M} \left(Y \hat{Y}'^T+\hat{Y}'Y^T\right)\,.
\end{equation}
Then, $\hat{Y}'$ and $Y$ are completely determined~\cite{Gavela:2009cd}, up to an overall scale, by the neutrino masses and mixings. 
On the other hand, the impact of $Y$, $\mu$ and $\mu'$ in the mass splitting $\Delta M$ of the pseudo-Dirac pair is very different, which implies different relations between light neutrino masses and heavy mass splittings for each scenario. Furthermore, while having simultaneously ISS and LSS terms is certainly possible, high-energy realizations based on symmetry arguments to obtain the necessary $L$-conserving texture tend to lead to only one source of $L$ breaking (see for instance~\cite{Malinsky:2005bi,Bazzocchi:2010dt,Khalil:2010iu,Okada:2012np,Ma:2014qra,Wang:2015saa,DeRomeri:2017oxa,Antusch:2017tud,Das:2019pua,Mandal:2021acg,Fernandez-Martinez:2021ypo,Abada:2021yot,Mondal:2021vou,Arias-Aragon:2022ats}). Thus, in the following will consider separately each of the realizations.  

In the case of the LSS, $\Delta M$ is of the same order of the light neutrino masses and in the case of a single HNL pair it is fixed by the light neutrino mass difference~\cite{Shaposhnikov:2008pf,Antusch:2017ebe,Drewes:2019byd} 
\begin{equation}  \label{DeltaM_LSS}
\Delta M^{\rm LSS} = \Delta m_\nu.
\end{equation}
On the other hand, the ISS contributions $\mu$ and $\mu'$ directly generate a HNL mass splitting, $\Delta M = \mu + \mu'$, and comparing this with their respective contributions to $m_\nu$ in this minimal scenario we can write:
\begin{align}
\Delta M^{\rm ISS} &=\frac{\Delta m_\nu}{\theta^2}\,, \label{DeltaM_ISS}\\
\Delta M^{\text{loop-ISS}} &=  \frac{16 \pi^2 \EWvev^2}{\left(M^2_H+3M^2_Z \right)} \frac{\Delta m_\nu}{\theta^2}\,.  \label{DeltaM_loopISS}
\end{align}
From these equations, it is then clear that we expect very different HNL mass splittings in each realization of the low-scale seesaws.
Roughly speaking, $\Delta M$ is of the same order of light neutrino masses in the LSS, while it is enhanced by two inverse powers of the small mixings in the ISS, and even further in the loop-ISS by an additional loop factor.
Thus, in principle, we should expect $\Delta M^{\text{loop-ISS}} > \Delta M^{\rm ISS} > \Delta M^{\rm LSS} $.

In order to illustrate the importance of this difference between low-scale seesaws, we follow Ref.~\cite{Drewes:2019byd} and consider the implications for the observability of lepton number violating (LNV) processes.
First, let us recall that, when mediated by a single Majorana HNL, the same rates for LNV and lepton number conserving (LNC) processes are expected.
As in the low-scale seesaws $L$ is only softly broken, the naive expectation is that LNV processes should also be very suppressed, which in practice happens due to the destructive contributions of each member of the pseudo-Dirac HNL pairs\footnote{Moreover, this is accompanied by a constructive interference for the LNC processes, see for instance~\cite{Abada:2022wvh}.}. 
Nevertheless, when the LNV process is mediated by HNL produced on-shell, it is possible that oscillations between the members of pseudo-Dirac pair happen, which could prevent the cancellation suppressing LNV signals in low-scale seesaws.

Whether HNL oscillations are effective is controlled by the relative size of two scales: the mass splitting between them, which controls their oscillation frequency, and their lifetimes (or equivalently the decay width $\Gamma_N$).
If $\Delta M \ll \Gamma_N$, HNLs will decay before they have time to oscillate, so the destructive interference between the HNLs is efficient and LNV processes are suppressed.
On the other hand, if $\Delta M \gg \Gamma_N$, HNLs oscillate many times before decaying, their effects are averaged out and similar rates for LNV and LNC processes are expected, as in the single Majorana HNL case. 
The coherence needed in order to have an effective destructive interference for LNV processes is lost. 

\begin{figure}[t!]
\begin{center}
\includegraphics[width=.9\textwidth]{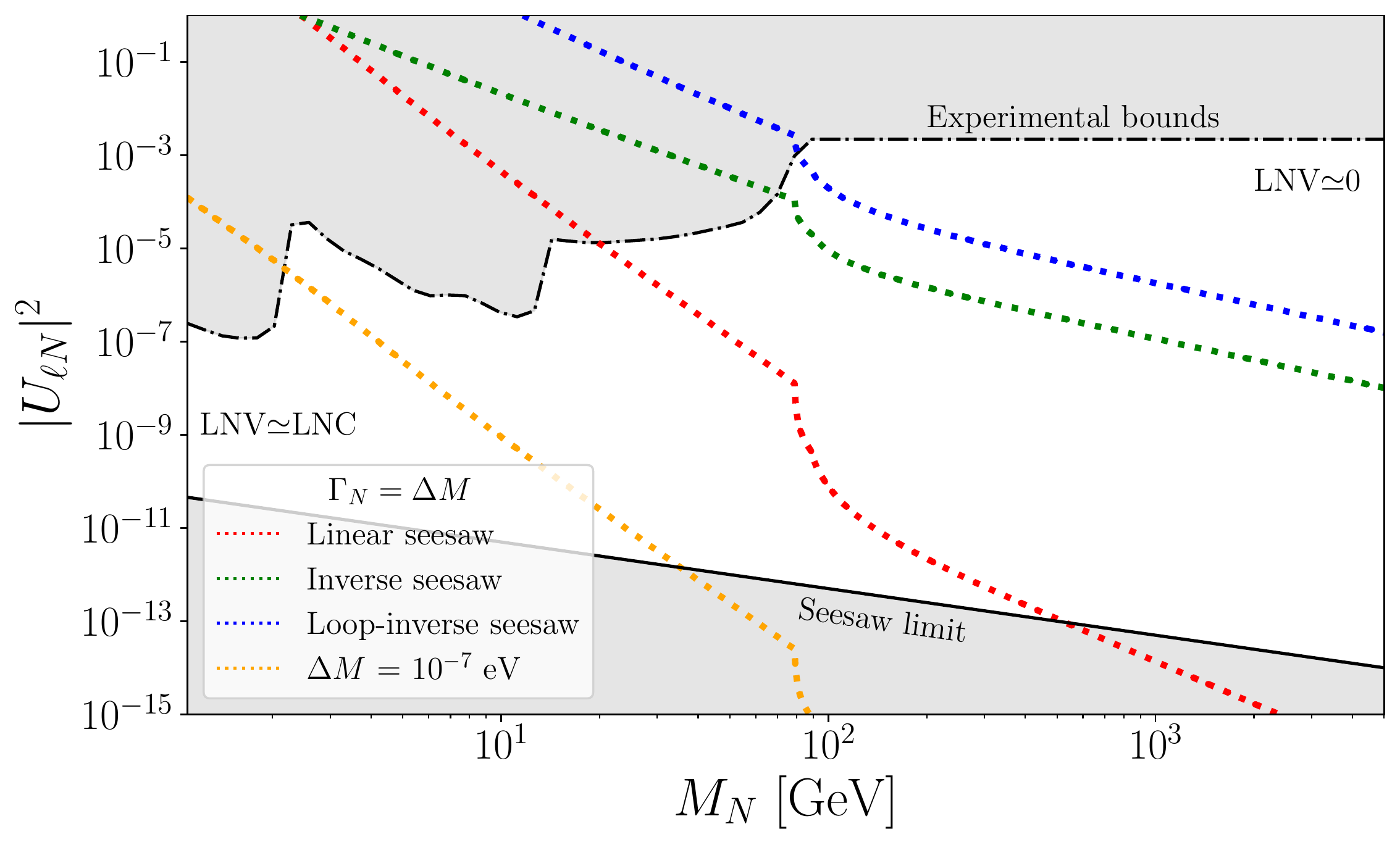}
\caption{Contourlines for the condition $\Gamma_N=\Delta M$ in each low-scale seesaw. Above (below) these lines we expect LNV processes to be (un)suppressed with respect to the LNC ones.
Also shown the line corresponding to the tiny $\Delta M$ shown to work for explaining DM via a sterile neutrino produced through the Shi-Fuller mechanism.
As a reference, we show the laboratory bounds for $U_{eN}$~\cite{Bolton:2019pcu,MatheusRepository} (gray shadowed area in the upper part), and the seesaw limit (solid black line) below which the HNLs generate too small active neutrino masses.
}\label{FeelingPlot}
\end{center}
\end{figure}

The HNL decay width, being a LNC observable, is the same in every low-scale seesaw realization and is determined by its mass and mixings to active neutrinos (see for instance Ref.~\cite{Coloma:2020lgy}).
The HNL mass splitting, on the other hand, is generated only via the $L$ breaking terms and thus is different in each low-scale seesaw, as can be explicitly seen from Eqs.~\eqref{DeltaM_LSS}-\eqref{DeltaM_loopISS}.
Consequently, the condition $\Delta M = \Gamma_N$ that separates the regime where LNV processes are suppressed or not will be different for each low-scale seesaw.
This is quantified in Fig.~\ref{FeelingPlot}, where we show the $\Delta M = \Gamma_N$ condition for the LSS, ISS and loop-ISS\footnote{For the LSS we assume NO and for the (loop-)ISS a reference scale of $m_\nu=\sqrt{\Delta m^2_{\text{atm}}}\approx 0.05$~eV.}
Above each of these lines, the corresponding model predicts a suppression for the LNV processes.
Bellow them, coherence is lost and LNV processes are no longer suppressed. 
We see that the regimes with suppressed and unsuppressed LNV are clearly different for each of the low-scale seesaws, which provides a way to experimentally distinguish them.

Figure~\ref{FeelingPlot} also shows the  mass splitting of $\Delta M \sim 10^{-7}$~eV~, 
which has been shown in Ref.~\cite{Ghiglieri:2020ulj} to generate enough lepton asymmetry so that the correct sterile neutrino dark matter is obtained via the Shi-Fuller mechanism. 
While the position of this line falls into the region of interest of many experiments (such as DUNE~\cite{Coloma:2020lgy}), generating such a small $\Delta M$ with just a single HNL pair is impossible with only one source of $L$ violation as in the LSS or ISS. Instead a cancellation of many orders of magnitude between the different contributions in Eqs.~\eqref{DeltaM_LSS}-\eqref{DeltaM_loopISS} would be required. In practice, the LSS term would provide the dominant contribution to the light neutrino masses $m_\nu$, leading to a contribution of similar order to $\Delta M$ via Eqs.~\eqref{DeltaM_LSS}, too large by 5 or 6 orders of magnitude. Then an opposite contribution to $\Delta M$ through the ISS parameters $\mu$ or $\mu'$ would be needed to bring it to the reference value of $O(10^{-7})$~eV while having a negligible impact in $m_\nu$, as long as $\theta$ is small enough. This very fine-tuned cancellation between parameters of a priori different origins (a Yukawa coupling versus a Majorana mass term) seems difficult to justify in a natural way and, moreover, it is highly unstable under higher order corrections.

In the rest of the paper, we will go beyond this minimal 2-HNL scenario, including additional HNLs in order to investigate if such small values of $\Delta M$ are achievable when only the LSS or ISS terms are considered separately (and therefore no cancellations between them can be present). We will also explore how the sharp predictions of the boundaries between LNC and LNV behaviours depicted in Figure~\ref{FeelingPlot} may be modified in these non-minimal scenarios. Thus, for the rest of the paper, we will consider only the pure LSS or pure ISS cases with several HNL pairs, which also correspond to most high-energy realizations of these scenarios. Indeed, when symmetry arguments to obtain the necessary $L$-conserving texture or a dynamical origin of the $L$ symmetry breaking are considered, only one source of $L$ breaking is typically present.

\section{Linear seesaw}
\label{Sec:LSS}

We start considering a pure LSS scenario with $n$ pairs of HNLs, with a neutrino mass matrix given by
\begin{equation}\label{MneuLSS}
    \mathcal{M}_\nu=
    \begin{pmatrix}
        0& Y\EWvev/\sqrt{2}&  Y'\EWvev/\sqrt{2}\\
         Y^T \EWvev/\sqrt{2}&0&M\\
        Y'^T\EWvev/\sqrt{2} &M&0
    \end{pmatrix}.
\end{equation}
Here, the Yukawa couplings $Y$ and $Y'$ are generic complex $3\times n$ matrices, with $Y'$ containing the small $L$ breaking parameters, and $M$ is a heavy $n\times n$ mass matrix that we can take as real and diagonal without any loss of generality.
It is possible to map this mass matrix to the canonical type-I seesaw in Eq~\eqref{type1SSmassmatrix} identifying:
\begin{equation}
    m_D=\left(Y\hspace{0.2cm}Y'\right)\EWvev/\sqrt{2}\,,\hspace{1cm}M_M=\begin{pmatrix}
0&M\\M&0   
\end{pmatrix}\,,
\end{equation}
so the light neutrino mass matrix reads:
\begin{equation}
    m_\nu=-\dfrac{\EWvev^2}{2}\left(Y M^{-1}Y'^T+Y'M^{-1}Y^T\right).
    \label{LSS:mnu}
\end{equation}

On the other hand, in the LSS the HNL pairs are degenerate at leading order, since $M_M$ has eigenvalues that are exactly degenerate in absolute value.
The small mass splitting is generated from the $O\left(\Theta^2 M_M\right)$ terms in Eq.~\eqref{type1SSMN}, which are of the same order as the contribution to the light neutrino masses, as we already saw explicitly in Eq.~\eqref{DeltaM_LSS} for a single HNL pair.
In presence of more pseudo-Dirac pairs, the model has more free parameters and this simple relation does not need to hold anymore. 
Nevertheless, it is still possible to connect the HNL mass splittings with light neutrino masses by considering trace relations.

In the LSS, the neutrino mass matrix $\mathcal M_\nu$ is traceless. 
Then, using the block-diagonalization in Eq.~\eqref{type1SSBlockDiagonalisation}, we have
\begin{equation}
	0=\Tr{\mathcal{M}_\nu}=\Tr{V^* \mathcal{M}_\nu^{\text{block}}V^\dagger}\,,
\end{equation}
and using Eq.~\eqref{type1SSVmatrix} to evaluate  the r.h.s.~explicitly, we get
\begin{equation}\label{Trace1}
    \Tr{V^* \mathcal{M}_\nu^{\text{block}}V^\dagger}\simeq\Tr{m_\nu}+\Tr{\Theta^T\Theta^* M_M}+2i\Tr{\Theta^T\imag{\Theta}M_M}=0\,.
\end{equation}
The second term in this equation can be related to the HNL mass splittings. 
In order to do so, we first recall that the matrix $M_M$ is traceless, so using Eq.~\eqref{type1SSMN} we can identify $\Tr{\Theta^T\Theta^* M_M}=\Tr{M_N}$.
Second, we notice that at leading order the heavy $M_N$ matrix has degenerate eigenvalues by pairs but with opposite sign. 
Thus, when we include the small perturbation induced by the second term in Eq.~\eqref{type1SSMN} and take the trace, only the small mass splittings will survive, {\it i.e},
\begin{equation}
    \Tr{M_N}=\sum_{i=1}^n \Delta M_i\,.
\end{equation}
Therefore, we can rewrite Eq.~\eqref{Trace1} as
\begin{equation}
    \sum_{i=1}^n \Delta M_i+\Tr{m_\nu}=-2i\Tr{\Theta^T\imag{\Theta}M_M}\,.
    \label{LSS:MasterRelation}
\end{equation}
This equation applies to a LSS with any number of pairs and represents the generalization of the simple $\Delta M = \Delta m_\nu$ relation in the minimal scenario with only one pair of HNLs. 
Notice however that in presence of several HNL pairs, it is possible to have some pairs with very small mass splittings, as long as the rest of the pairs have large enough $\Delta M$ to satisfy this equation.
Furthermore, it is in principle posible to have cancellations between the two traces in Eq.~\eqref{LSS:MasterRelation}, so every HNL pair could have a very small mass splitting while still reproducing the correct light neutrino masses and mixings. 
We will show an explicit example of such configuration later in our numerical analysis.

The r.h.s.~of Eq.~\eqref{Trace1} originates from the fact that the block-diagonalisation of Eq.~\eqref{type1SSVmatrix} is not trace preserving if there are complex phases in the Yukawa couplings ({\it i.e.}~$V^T V\neq 1$).
In the case of real parameters, the r.h.s is zero and we have a more direct relation between HNL mass splittings and light neutrino masses\footnote{Notice however that Tr{$\{m_\nu\}$} is not necessarily equal to the sum of neutrino masses, since there are still phases involved in its diagonalization. We will discuss it in more detail later.}.
We explore these two hypotheses of real and complex parameters in the following, focusing on the case of having two pairs of HNLs. 
While being the minimal extension with respect to the simplest LSS model, we will see that it already has enough freedom to cover scenarios with HNL mass splittings much smaller or bigger than the naive expectations.

\subsection{Next-to-minimal LSS}
We will now focus on studying the LSS with two pseudo-Dirac pairs, which is the simplest extension of the minimal LSS with one pair. 
Notice that, while not being the minimal setup to explain current oscillation data, it does contain the minimal amount of pairs that one needs to accommodate $3$ massive neutrinos.

From now on, we will use a similar notation to the one introduced in \cite{Gavela:2009cd}, where we parameterise the Yukawa couplings as unitary vectors in flavor space times their modulus. Namely,
\begin{equation}
    Y=\left(y_1 \mathbf{u}_1\hspace{0.25cm}y_2\mathbf{u}_2\right),
    \hspace{0.75cm}
    Y'=\left(y'_1\mathbf{v}_1\hspace{0.25cm}y'_2\mathbf{v}_2\right),
\end{equation}
where $\mathbf{u}_i^\dagger \mathbf{u}^{\phantom \dagger}_i=1$ and $\mathbf{v}_i^\dagger \mathbf{v}^{\phantom \dagger}_i=1$ for $i=1,2$.
In this notation, the light neutrino mass matrix in Eq.~\eqref{LSS:mnu} takes the following form:
\begin{equation}\label{LSSmnu}
    m_\nu=-\epsilon_1\left(\mathbf{u}^{\phantom \dagger}_1\mathbf{v}_1^T+\mathbf{v}^{\phantom \dagger}_1\mathbf{u}_1^T\right)
    -\epsilon_2\left(\mathbf{u}^{\phantom \dagger}_2\mathbf{v}_2^T+\mathbf{v}^{\phantom \dagger}_2\mathbf{u}_2^T\right),
\end{equation}
where we have introduced the two light neutrino mass scales $\epsilon_i\equiv \frac{\EWvev^2 y_i y'_i}{2M_i}$, with the heavy scales $M_i$ given by the entries of the diagonal matrix $M$.

On the other hand, using Eq.~\eqref{type1SSMN}, the HNL mass splittings can be computed in terms of the model parameters in this parametrisation:
\begin{equation}
    \Delta M_i\approx 2\epsilon_i \rho_i\,,
    \label{2LSS:SplittingsExpression}
\end{equation}
where we have defined $\rho_i=\mathbf{u}_i^\dagger \mathbf{v}^{\phantom\dagger}_i$ as the scalar product of the two unitary Yukawa directions of each pair, which can be always considered real independently of the Yukawa couplings being complex\footnote{This is due to the fact that we can always rephase $N$ and $N'$ such that $N\rightarrow e^{-i\varphi/2} N\hspace{0.1cm},\hspace{0.1cm}N'\rightarrow e^{-i\varphi/2}N'$, which in turn induces a rephasing of $\rho\rightarrow e^{i\varphi}\rho$ without introducing phases in any quantity that was previously real. Using this freedom we can rephase away any complex phase in $\rho$, thus keeping it real.}.

Once more, we see explicitly that the HNL mass splittings in the LSS are controlled by the light neutrino mass scales $\epsilon_i$, although they are additionally weighted by the scalar product $\rho_i$. 
In the minimal LSS model with 1 HNL pair, there is only one $\rho$ and it is fixed by light neutrino masses~\cite{Gavela:2009cd}, so we recover Eq.~\eqref{DeltaM_ISS}.
Now, in our next-to-minimal LSS with 2 pairs, the relation is more involved, however we can still relate the size of the HNL mass splittings to light neutrino masses by means of Eq.~\eqref{LSS:MasterRelation}.
In the following, we analyse this relation in detail assuming real parameters first, and generalising to complex ones afterwards.

\subsection*{Real Yukawa case}
 If the Yukawas are assumed to be real, a rather simple relation can be found between the pseudo-Dirac mass splittings and the light neutrino masses, since the r.h.s~of Eq.~\eqref{LSS:MasterRelation} vanishes. This is just a consequence of the fact that, when the Yukawa couplings are real, the block-diagonalisation matrix in Eq.~\eqref{type1SSVmatrix} is orthogonal and the trace is preserved.
 Thus, Eq.~\eqref{LSS:MasterRelation} reduces to
 \begin{equation}
     \Delta M_1+\Delta M_2=-\Tr{m_\nu}\,.
     \label{LSSReal:MasterRelation}
 \end{equation}
 
 Now, since the PMNS matrix in this case is orthogonal ($\delta_{CP}=0$), the diagonalisation of the light sector is also trace-preserving and the trace of $m_\nu$ will be the sum of its eigenvalues. 
 However, this does not mean that it will be equal to the sum of neutrino masses due to the fact that all the eigenvalues of $m_\nu$ must be real, but not necessarily positive\footnote{Notice that neutrino masses are obtained from the square root of the positive eigenvalues of $m_\nu^\dagger m_\nu$ and that a negative sign in an eigenvalue of $m_\nu$ only amounts to a Majorana phase. However, to relate LNV quantities such as the pseudo-Dirac pair mass splitting with the measured neutrino masses through Eq.~\eqref{LSS:MasterRelation}, these phases are relevant}. 
 Indeed, the matrix in Eq.~\eqref{LSSmnu} will always have one eigenvalue with opposite sign to the other two\footnote{Each matrix $\hat{O}=\mathbf{u}\mathbf{v}^T+\mathbf{v}\mathbf{u}^T$ has two non-zero eigenvalues $\lambda_\pm=\pm1+\rho$ where $\abs{\rho}=\abs{\mathbf{u}^\dagger \mathbf{v}}<1$ and it can be shown that the sum of two such matrices, and therefore $m_\nu$, generally has three non-vanishing eigenvalues with one of them having opposite sign with respect to the other two.}. The choice of which mass eigenvalue has flipped its sign will depend on the particular configuration of the Yukawa couplings of the model, leading to three different values for $\Tr{m_\nu}$ (up to a global sign):
 \begin{equation}
     \abs{\Tr{m_\nu}}=\left\lbrace\begin{array}{c} \abs{m_2+m_3-m_1}\equiv t_1\,, \\ \abs{m_1-m_2+m_3}\equiv t_2\,, \\ \abs{m_1+m_2-m_3}\equiv t_3\,, \end{array}\right.
     \label{2LSS:TraceValuesReal}
 \end{equation}
 with $m_i$ the physical masses of light neutrinos, given by the absolute value of the eigenvalues of $m_\nu$.
 
 As for the l.h.s.~of Eq.~\eqref{LSSReal:MasterRelation}, one also has to take into account possible signs, since our physically interesting quantity is the absolute value of the mass splittings, and not their sign. 
 Following Eq.~\eqref{2LSS:SplittingsExpression}, its sign depends on the sign of $\rho_i$ which, again, depends on the particular configuration of the Yukawa couplings.
 Therefore, if one marginalises over all the possible sign combinations both for $\Delta M_i$ and $\Tr{m_\nu}$, we arrive at the expressions for curves in which $\abs{\Delta M_i}$ can lie
 \begin{equation}
	\abs{\Delta M_1}+\abs{\Delta M_2}=t_{1,2,3}\,,\hspace{.8cm}\abs{\Delta M_1}-\abs{\Delta M_2}=t_{1,2,3}\,,\hspace{.8cm}\abs{\Delta M_2}-\abs{\Delta M_1}=t_{1,2,3}\,.
	\label{2LSS:TraceRelationsReal}
\end{equation}

\begin{figure}[t!]
    \centering
    \includegraphics[width=\textwidth]{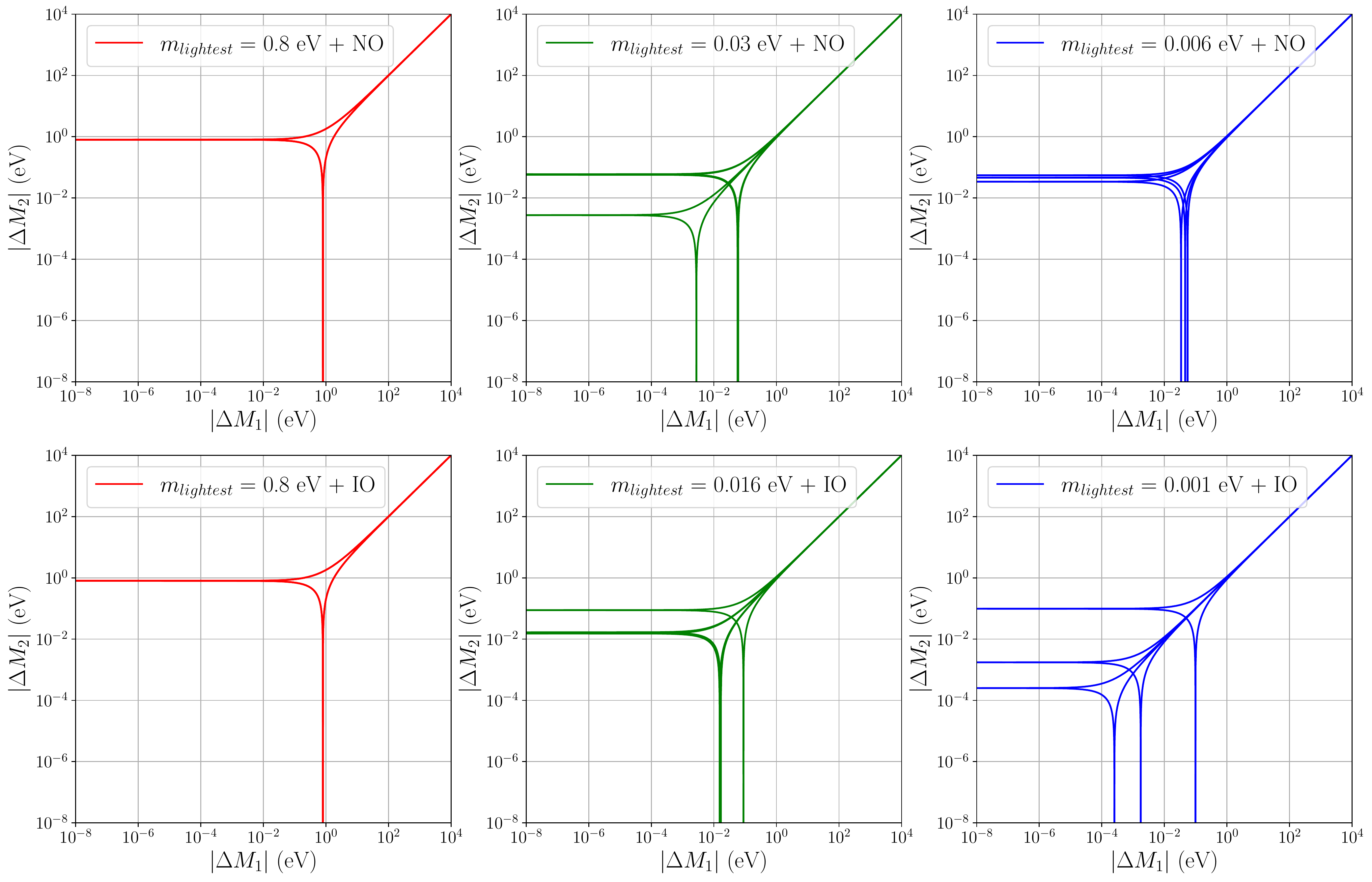}
    \caption{Correlation curves along which the tree-level HNL mass splittings must lie in order to accommodate oscillation data in the next-to-minimal LSS with real Yukawas. 
    Upper (lower) panels are for NO (IO), each for a different hypothesis of lightest neutrino mass: KATRIN upper limit (left panels in red), the $\Lambda$CDM cosmological upper bound of $\sum m_\nu=0.12$~eV~\cite{Planck:2018vyg} (central panels, green), and an {\it ad-hoc} scale that makes the three triangular curves more distinct and visible (right panels, blue), which is more pronounced in the case of the IO.
    }
    \label{2LSS:LineasTeoricas}
\end{figure}

Figure~\ref{2LSS:LineasTeoricas} shows the shape of these lines in the mass splittings space, for different choices of the lightest neutrino mass and neutrino mass ordering, be it Normal Ordering (NO) or Inverted Ordering (IO).
In this logarithmic scale, we find that the HNL mass splitting must lie along one of these three triangular-shaped lines centered around $(t_i,t_i)$.
For $\mlightest=0.8$~eV, the maximum allowed value by KATRIN~\cite{KATRIN:2021uub}, light neutrinos are almost degenerate, $t_1\simeq t_2\simeq t_3\simeq 0.8$~eV, and the three lines collapse to a single one for both NO and IO.
As we lower the value of $\mlightest$ the three traces $t_i$ take different values, so the correlation curves separate depending on the ordering. 
Indeed, we see that for the chosen values of $\mlightest$, in the case of NO two of the traces are always almost degenerate, whereas for IO there are values of $\mlightest$ for which the traces have substantially different values.

From these results, it is clear that the mass splitting of one of the pairs can vanish as long as the other mass splitting is equal to one of the traces $t_{1,2,3}$. 
Getting to a scenario where both $\Delta M_i$ could vanish at the same time seems difficult in this case, although there is a particular value of $\mlightest$, different for each mass ordering, for which one of the traces vanishes, allowing both HNL mass splittings to be arbitrarily small.
On the opposite direction, we also find solutions for which the HNL mass splittings, both of them actually, can be enhanced considerably with respect to their naive expectations of $\mathcal O(m_\nu)$.

\begin{figure}[t!]
    \centering
    \includegraphics[width=1\textwidth]{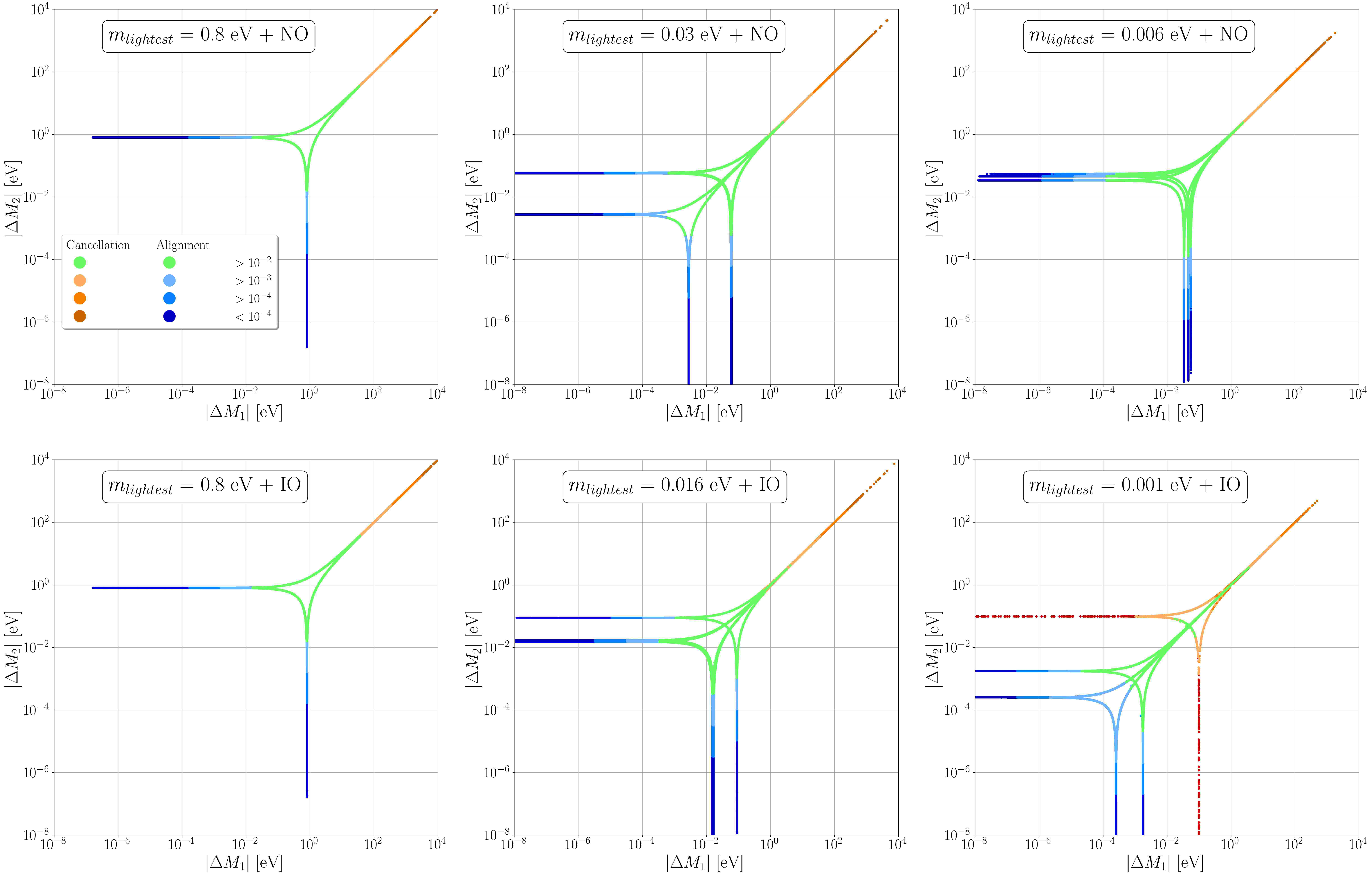}
    \caption{Numerical scan of the next-to-minimal LSS with real Yukawa couplings, showing the mass splittings of the two pairs of HNLs after imposing agreement with oscillation data.
    Blue points correspond to configurations where some Yukawas are almost perfectly orthogonal, while orange ones to those where large cancellations between parameters are happening.
    Green points do not fall into any of these categories. Furthermore, in the last plot we have shown in dark red the directions along which the parameter configurations exhibit both large cancellations and almost orthogonal Yukawas. 
    Further details are given in the text. 
    }
    \label{2LSS:MultiDegradedScatters}
\end{figure}

In order to explore better these correlations and to understand under which conditions we can get to the extreme values of the HNL mass splittings, we perform a numerical scan of the parameter space of this next-to-minimal LSS compatible with current oscillation data.

The details on how we perform this scan are given in Appendix \ref{Appendix:LSSRealScan}, while the results are shown in Fig.~\ref{2LSS:MultiDegradedScatters}.

We notice that the numerical results follow the analytical description in Eq.~\eqref{2LSS:TraceRelationsReal} very closely.
Nevertheless, this new figure allows to distinguish three regions with different behaviours: firstly, a regime shown in green in which the two mass splittings are as naively expected, both of the order of the traces $t_i$; secondly, a direction (in orange) in which the two mass splittings become larger and their values are almost identical; and, finally, directions (in blue) along which one of the mass splittings becomes arbitrarily small, as the other one saturates to the values of one of the three traces $t_{1,2,3}$.

These regimes can be understood in terms of the parameters of our model and how they translate into the mass splittings (namely through Eq.~\eqref{2LSS:SplittingsExpression}). 
The orange regime corresponds to the two light neutrino mass scales $\epsilon_i$ becoming larger than the physical neutrino mass scales. Thus, in order to reproduce the correct neutrino masses, {\it i.e.}~avoiding too large light neutrino masses, the contributions from $\epsilon_1$ and $\epsilon_2$ need to cancel each other precisely. Hence, both scales become very similar, giving rise to very similar mass splittings. As such, increasing $\Delta M$ in this regime implies a very precise cancellation between the two contributions, signalling either some symmetry relating them or a significant fine-tuning of the model.
In order to quantify the amount of cancellation between the two mass scales, we
define the following quantity
\begin{equation}
    c_\epsilon=\abs{\dfrac{\epsilon_1-\epsilon_2}{\epsilon_1+\epsilon_2}}\,,
\end{equation}
and color-code the points in our parameter scan with different shades of orange. In particular Fig.~\ref{2LSS:MultiDegradedScatters} depicts points with $c_\epsilon < 10^{-2}, 10^{-3}$ or $10^{-4}$ (from lighter to darker orange), showing that enhancing the HNL mass splitting with respect to their naive values $t_{i}$ requires increasingly more precise cancellations between the, a priori, independent parameters $\epsilon_1$ and $\epsilon_2$.

On the other hand, the regime in blue corresponds to points with one of the mass splittings being very small. According to Eq.~\eqref{2LSS:SplittingsExpression}, this situation corresponds to either one of the $\epsilon_i$ or one of the $\rho_i$ becoming very small. 
However, there is a limit on how small the scales $\epsilon_i$ can be in order to reproduce the correct neutrino masses, {\it i.e.}, if the contribution of one of the pairs is too small, one of the neutrinos will not acquire enough mass. 
Thus, this regime actually corresponds to one of the $\rho$ going to $0$. 
In other words, one of the pseudo-Dirac pairs may have a very small or even zero splitting if its Yukawa directions are orthogonal. Conversely, the other pair must have a splitting set by one of the quantities $t_{1,2,3}$.

As before, it is interesting to quantify how orthogonal the Yukawa vectors need to be in order to allow for such small mass splittings. 
We do this by color-coding the values of the scalar products $\rho_1$ and $\rho_2$, which measure the alignment between the Yukawa vectors.
In particular, we show in blue those points with one of the $\rho_i<10^{-2}, 10^{-3}$ or $10^{-4}$ (from lighter to darker blue). 
Notice that, in order to achieve the reference $\Delta M\sim 10^{-7}$~eV for the Shi-Fuller mechanism for DM neutrino production, a HNL pair whose Yukawas are orthogonal to a precision of roughly $3$ to $4$ orders of magnitude would be necessary. Interestingly, there are models based on symmetry arguments where some Yukawas are orthogonal, see for instance~\cite{Altarelli:1999dg,Antusch:2004gf,Bjorkeroth:2014vha}.

Finally, green points in Fig.~\ref{2LSS:MultiDegradedScatters} correspond to intermediate choices of the parameters in which there are no significant cancellations or orthogonal Yukawa couplings and thus, both mass splittings lie close to the naive expectations. 
Nevertheless, and before moving on to complex Yukawa couplings, we would like to stress again that the color coding in Fig.~\ref{2LSS:MultiDegradedScatters} is not meant to exclude scenarios with enhanced or suppressed HNL mass splittings, but rather to quantify the ``price to pay'', either through fine-tuning or some additional symmetry argument, in order to move along the different directions.

Moreover, notice that even if the Yukawa couplings of one of the pairs are perfectly orthogonal and there is a perfect cancellation of $\Delta M$ at tree level, this cancellation is not stable under higher order corrections in absence of any additional symmetries.

\subsection*{Complex Yukawa case}
We now consider the generalization of the previous discussion when switching on the possible CP phases so as to also allow for non-zero values of the PMNS phase $\delta_{\rm CP}$ or the necessary CP-violation sources in order to have a successful leptogenesis mechanism. 

The crucial difference between having real and complex Yukawas is that now the transformation that diagonalises the mass matrix is not trace preserving. 
Namely, the r.h.s~of Eq.~\eqref{LSS:MasterRelation} does not vanish, the traces of the heavy and light sectors are not equal, and thus the HNL mass splittings satisfy
\begin{equation}
    \Delta M_1+\Delta M_2=-\Tr{m_\nu}-2i\Tr{\Theta^T\imag{\Theta}M_M}.
    \label{ComplexCase_MasterRelation}
\end{equation}
Moreover, both the Majorana and Dirac phases of the light neutrino sector can have arbitrary values.
Then, $\Tr{m_\nu}$ is not always given by the sum (up to relative signs) of the light neutrino masses, but it takes the more general form of
\begin{equation}
    \Tr{m_\nu}=\Tr{U'^* \mdiag_\nu U'^\dagger}=\sum_{\alpha,i}\left(U'^*_{\alpha i}\right)^2m_i\,,
\end{equation}
where $U'=\UPMNS\cdot U_M$ is the unitary matrix that diagonalises $m_\nu$ in Eq.~\eqref{Eq:fullmatrix}, including the possible Majorana phases. 

In analogy with the effective mass for the neutrinoless double decay process, this quantity can be seen as the sum of effective Majorana masses of the three lepton flavours
\begin{equation}
    \Tr{m_\nu}=\sum_{\alpha=e,\mu,\tau} m_{\alpha\alpha}\,.
\end{equation}
It turns out that, except for very small values of $\mlightest$, this quantity is not bounded from below, which is in stark contrast with the case without phases in which $\Tr{m_\nu}$ had only $3$ possible values (cf.~Eq.~\eqref{2LSS:TraceValuesReal}). Furthermore, a lower bound for the second term in the r.h.s.~of Eq.~\eqref{ComplexCase_MasterRelation} cannot be set only from light sector quantities, as it contains mixings and phases belonging to the heavy sector. Thus, in general, this term may also vanish.
Therefore, it is possible to make the r.h.s.~of Eq.~\eqref{ComplexCase_MasterRelation} vanish for certain Yukawa configurations, which would imply vanishing mass splittings (at tree level) for both pseudo-Dirac pairs.
Notice that this configuration was not possible in absence of complex phases unless $\mlightest$ had the precise value so as to cancel one of the $t_i$ in Eq.~\eqref{2LSS:TraceValuesReal}.

In order to showcase the existence of such configurations in which exact cancellations of both mass splittings at tree level are present, we consider a particular case that can be treated analytically. Notice that, in order for $\Delta M_i$ to vanish, $\rho_i=0$ is required, {\it i.e.} the two Yukawa directions of each pair ($\mathbf{u}_i$ and $\mathbf{v}_i$) need to be orthogonal. 
We consider the following setup:
\begin{align}\label{ISSexample}
     \mathbf{u}_1^\dagger \mathbf{v}^{\phantom \dagger}_1&=\rho_1=0\,, &
     \mathbf{u}_2^\dagger \mathbf{v}^{\phantom \dagger}_2&=\rho_2=0\,,  \nonumber\\
     \mathbf{u}_1^\dagger \mathbf{u}^{\phantom \dagger}_2 &=e^{i\alpha}/2\,, &\mathbf{v}_1^\dagger \mathbf{v}_2^{\phantom \dagger}&=e^{i\beta}/2\,,\nonumber\\
     \mathbf{u}_1^\dagger \mathbf{v}^{\phantom \dagger}_2&=e^{i\gamma}/2\,, &\mathbf{u}_2^\dagger \mathbf{v}^{\phantom \dagger}_1&=e^{i\delta}/2\,.
\end{align}
This configuration corresponds to the $4$ Yukawas disposed along the edges of a square pyramid with equilateral triangle sides, and (by construction) it yields $\Delta M_i=0$ at tree level, since $\rho_1=\rho_2=0$. Furthermore, even though in the remaining $4$ scalar products one can consider $4$ different complex phases $(\alpha,\beta,\gamma,\delta)$, they are not independent and in fact only $2$ phases are physical: the phases in $\mathbf{v}_1^\dagger \mathbf{v}^{\phantom \dagger}_2$ and $\mathbf{u}_2^\dagger \mathbf{v}^{\phantom \dagger}_1$ can be set to $0$ due to the freedom of rephasing each of the two pseudo-Dirac pairs separately. 

The moduli of the mass eigenvalues of this particular configuration are:
\begin{equation}
    m_i=\left\{\begin{array}{l}
         \epsilon_1\,, \\
         \epsilon_2\,,\\
         \sqrt{\epsilon_1^2+\epsilon_2^2+2\epsilon_1\epsilon_2\cos\xi}\,,
    \end{array}\right.
    \label{eq:complexcase}
\end{equation}
where $\xi$ is the only phase combination that contributes to neutrino masses (the other combination appears in the form of Majorana phases). This equation shows the impact of allowing for complex Yukawa couplings: in the real limit ($\xi=0,\pi$), the three mass eigenvalues are correlated, since one neutrino mass would be fixed given the other two. In this context, if we want to reproduce the correct mass splittings for light neutrinos while still having $\Delta M_i=0$, there is only one particular value of $\mlightest$ which can do the job (which corresponds to the $\mlightest$ value that makes one of the three trace values $t_{1,2,3}$ in Eq.~\eqref{2LSS:TraceValuesReal} vanish). However, in presence of complex phases, there is additional freedom (an extra parameter) to accommodate the correct neutrino masses. Even though this is only a simplified example, it illustrates that, while the condition $\rho_1=\rho_2=0$ with real Yukawas is too restrictive so to explain neutrino masses, this is not the case in the presence of complex phases.

These solutions with suppressed HNL mass splittings are an example of how the presence of complex phases can distort the results found for the real case. Thus, it motivates to explore the more general behavior. 
Due to the additional degrees of freedom that the complex phases introduce, scanning the parameter space compatible with oscillation data in the same way as in the real case is not efficient. Indeed, there are now $7$ extra (physical) phases in the Yukawa couplings that need to be scanned over.
Consequently, in order to scan efficiently the complex Yukawa couplings that reproduce oscillation data, we will follow the master parametrisation presented in~\cite{Cordero-Carrion:2019qtu}, particularised for our next-to-minimal LSS.
The details of said parametrisation are shown in appendix~\ref{Appendix:AvelinoParametrisation}.

\begin{figure}[t!]
    \centering
    \includegraphics[width=\textwidth]{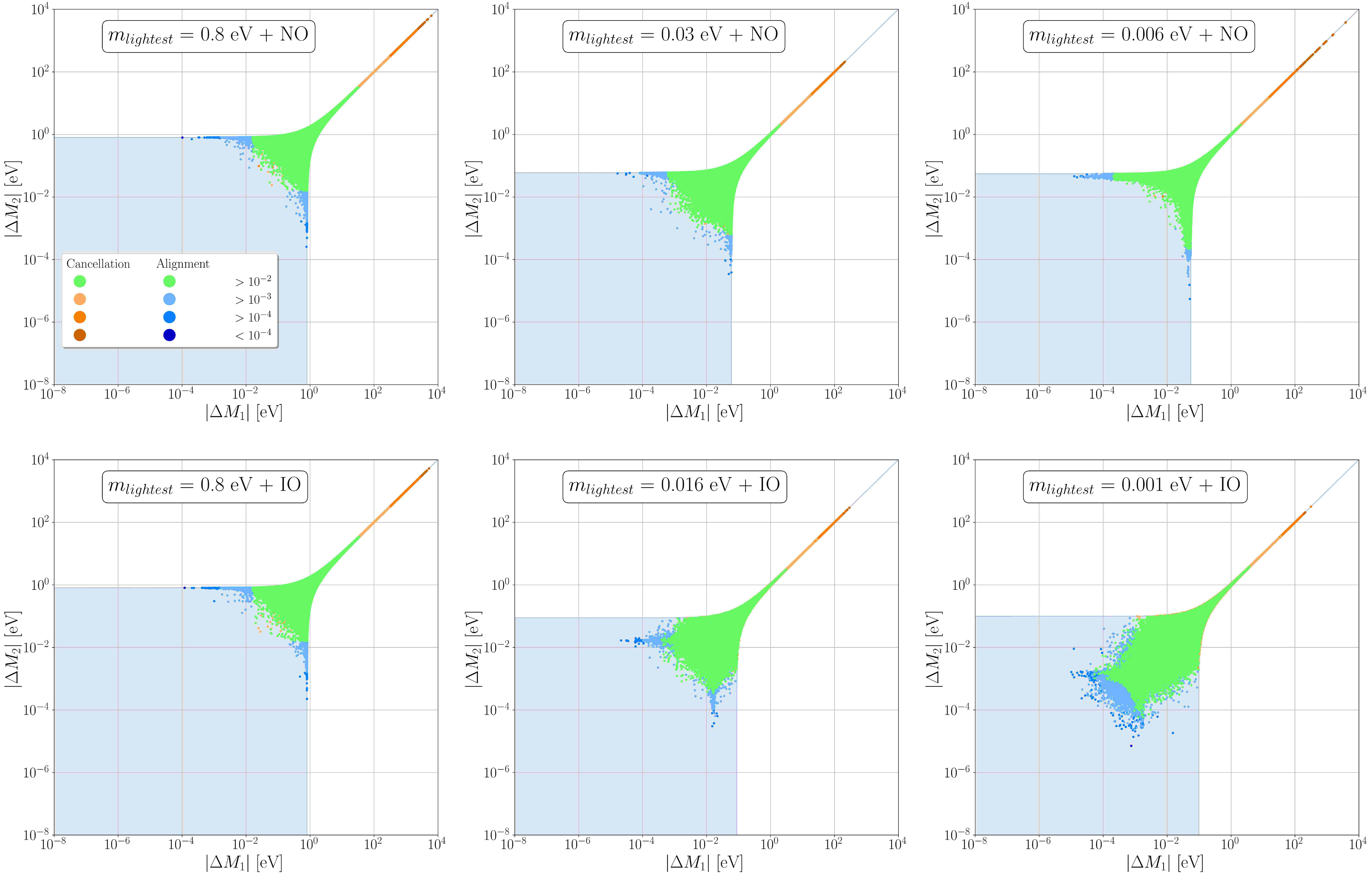}
    \caption{Same as Fig.~\ref{2LSS:MultiDegradedScatters}, but for complex Yukawa couplings.
    The regions highlighted in light blue represent the areas where solutions compatible with oscillation data exit, even if our numerical scan did not fully cover them.}
    \label{FIG:LSS_DegradedScatterComplex}
\end{figure}

The results of our scan are shown in Fig.~\ref{FIG:LSS_DegradedScatterComplex}, following the same color code as before.
We clearly see that the presence of complex phases loosens the strong correlations we found in the case of real Yukawa couplings, although the overall behaviour is similar and the qualitative discussion about the areas with large cancellations or very orthogonal Yukawa couplings still applies.
The biggest difference is that now we find solutions below the correlation lines in Fig.~\ref{2LSS:MultiDegradedScatters}, covering the area where the two mass splittings can be suppressed at the same time.

From the numerical solutions displayed in Fig.~\ref{FIG:LSS_DegradedScatterComplex}, it is not easy to conclude how small the $\Delta M_i$ could be, since the scan points cluster around the ``natural'' green region close to the light neutrino mass scales.
Indeed, as discussed above, the solutions for which much smaller values of the HNL mass splittings are possible correspond to very particular configurations, which are very difficult to realise through a random general scan unless they are enforced.
Nevertheless, we have already shown in Eq.~\eqref{ISSexample} an example where both $\Delta M_i$ can vanish and still reproduce oscillation data, which provides a ``proof-of-principle'' that all the region below the lines of the real case is now open\footnote{Indeed, we could go continuously from the $\Delta M_{1}=\Delta M_2=0$ point to the green area by adding small perturbations to the orthogonality conditions in Eq.~\eqref{ISSexample}.}. Furthermore, in the more easily tractable case with real Yukawas shown in Fig.~\ref{2LSS:MultiDegradedScatters}, we have shown and quantified the requirements on the parameters in order to have values of $\Delta M$ both above and below those lying in the green area. By comparing Figs.~\ref{2LSS:MultiDegradedScatters} and~\ref{FIG:LSS_DegradedScatterComplex} we also see that the main effect of adding complex phases is to allow smaller values that in the real case for the same level of cancellation in the model parameters.  

On the other hand, we do not find any solutions in our numerical scan above the values of the real case.
More precisely, we  find that all the points lie below one of these curves
\begin{equation}
    \big|\abs{\Delta M_1}-\abs{\Delta M_2}\big|\leq \max\left[t_1,t_2,t_3\right].
\end{equation}

Summing up, we conclude that in the complex next-to-minimal LSS it is possible to find solutions where both tree-level mass splittings are arbitrarily suppressed at the same time, while the scenario with real Yukawas provides an upper limit for the correlations between the two.
We display these regions, where we have shown solutions to exist, as blue shaded areas in Fig.~\ref{FIG:LSS_DegradedScatterComplex}.
The coloured points quantify the level of cancellation needed in order to reach the most extreme values in this space. While these areas are more difficult to populate in the most general setup with complex phases, we find that they align well with the more exhaustive results of the real case and expect a very similar behaviour. While these areas may be viewed as significantly fine-tuned, symmetry arguments could perhaps provide a rationale to expect those relations, particularly for the condition to obtain very small $\Delta M$ which requires the Yukawa directions in flavour space to be orthogonal~\cite{Altarelli:1999dg,Antusch:2004gf,Bjorkeroth:2014vha}.  

\section{Inverse seesaw}
\label{Sec:ISS}

In this section we consider a pure ISS realization, although we notice that our discussion also applies to the loop-ISS, with the corresponding HNL mass splittings enhanced by the loop factor as in Eq.~\eqref{DeltaM_loopISS}.
In general, assuming the presence of $n$ pairs of HNLs, the ISS neutrino mass matrix reads
\begin{equation}\label{MneuLSS}
    \mathcal{M}_\nu=
    \begin{pmatrix}
        0& Y\EWvev/\sqrt{2}&  0\\
         Y^T \EWvev/\sqrt{2}&0&M\\
        0 &M&\mu
    \end{pmatrix}\,,
\end{equation}
where, as before, the Yukawa $Y$ is a generic complex $3\times n$ matrix and $M$ is a heavy $n\times n$ mass matrix taken, without loss of generality, to be real and diagonal. 
The difference with respect to the LSS model is in the small $L$ breaking term, which is now in the complex symmetric $\mu$-matrix, whose entries are small compared with those of $M$. 

As in the previous section, we can map this matrix to the general type-I seesaw in Eq.~\eqref{type1SSmassmatrix}.
Then, we have
\begin{equation}
    m_D=\left(Y\hspace{0.2cm}0\right)\EWvev/\sqrt{2}\,,\hspace{1cm}M_M=\begin{pmatrix}
0&M\\M&\mu   
\end{pmatrix}\,,
\end{equation}
and the light neutrino mass matrix in Eq.~\eqref{type1SSmnu} reads
\begin{equation}\label{ISSmlight}
    m_\nu=\dfrac{\EWvev^2}{2}YM^{-1}\mu\hspace{0.1cm}M^{-1}Y^T\,.
\end{equation}
The heavy sector mass is still given by Eq.~\eqref{type1SSMN}, however $M_M$ is not traceless in the ISS, and therefore it already contains a contribution to the HNL mass splittings at leading order in the seesaw expansion.
More precisely, in the ISS the $M_M$ term contains the $\mu$-matrix, whose diagonal elements already contribute to the mass splittings, that is,
\begin{equation}\label{ISSdeltaMN}
\Delta M_{i} = \mu_{ii}\,.
\end{equation}
The $\mathcal O(\Theta^2 M_M)$ term in  Eq.~\eqref{type1SSMN} now introduces subleading contributions to the HNL mass splittings ({\it i.e.}~suppressed by the small mixing angles), which will be, in general, negligible. 

Comparing Eqs.~\eqref{ISSmlight} and \eqref{ISSdeltaMN}, we see that the HNL mass splittings and light neutrino masses are generated at different orders in the seesaw expansion and are not so directly correlated.
This is in sharp contrast with the case of the LSS in the previous section, where they were both of the same order.
In general, we could naively expect that each of the ISS pairs will have a splitting following Eq.~\eqref{DeltaM_ISS}, with a $\Delta M$ enhanced with respect to light neutrino masses by two inverse powers of the small mixing between the heavy neutrinos and the active flavour states. 
Nevertheless, Eq.~\eqref{ISSdeltaMN} already suggests that it is possible to find a scenario where $\mu$ is purely off-diagonal, so that $\Delta M_{i}=0$ at leading order, while still reproducing the correct neutrino masses and mixings due to the flavor structure of the Yukawa couplings in Eq.~\eqref{ISSmlight}.
Of course, in this limit with a vanishing leading contribution, the $\mathcal O(\Theta^2 M_M)$ term in  Eq.~\eqref{type1SSMN} will become relevant and generate a mass splitting with a similar dependence to the LSS case. However, this is still a suppression with respect to the naive expectations in the ISS. 

In order to explore the relation between Eqs.~\eqref{ISSmlight} and \eqref{ISSdeltaMN} in more detail, we consider the simplest ISS that can accommodate the current oscillation data~\cite{Abada:2014vea}. 
This model contains 2 HNL pairs, which give mass to only two light neutrinos, and, as we will discuss, already introduces enough freedom to lead to significant deviations from the naive ISS expectations for the HNL mass splittings.

\subsection{Minimal ISS}
In this minimal 2-pair scenario, the Yukawa matrix and the $\mu$-matrix take the form:
\begin{equation}
    Y=\left(y_1 \mathbf{u}_1\hspace{0.25cm}y_2\mathbf{u}_2\right)\,,
    \hspace{0.75cm}
    \mu=\begin{pmatrix}
        \mu_1&\mu_3\\\mu_3&\mu_2
    \end{pmatrix}\,,
\end{equation}
where, as in the LSS case, we have decomposed the Yukawa matrix in terms of two unitary vectors ($\mathbf{u}_1$ and $\mathbf{u}_2$) and their moduli ($y_1$ and $y_2$). This translates into a light neutrino mass matrix given by
\begin{equation}\label{mlightMinimalISS}
    m_\nu=\mutilde_1\mathbf{u}^{\phantom \dagger}_1\mathbf{u}_1^T+\mutilde_2\mathbf{u}^{\phantom \dagger}_2\mathbf{u}_2^T
    +\mutilde_3\left(\mathbf{u}^{\phantom \dagger}_1\mathbf{u}_2^T+\mathbf{u}^{\phantom \dagger}_2\mathbf{u}_1^T\right),
\end{equation}
where we have defined the three light neutrino mass scales as
\begin{equation}
    \mutilde_1=\mu_1 \dfrac{\EWvev^2y_1^2}{2M_1^2}=\mu_1U_1^2\,,\hspace{0.75cm}\mutilde_2=\mu_2 \dfrac{\EWvev^2y_2^2}{2M_2^2}=\mu_2U_2^2\,,\hspace{0.75cm}\mutilde_3=\mu_3\dfrac{\EWvev^2 y_1y_2}{2M_1M_2}=\mu_3U_1U_2\,,
\end{equation}
and the total squared mixing  of a given mass eigenstate to all active flavours as:
\begin{equation}
    U_i^2\equiv \sum_{\alpha=e,\mu,\tau} \abs{U_{\alpha N_i}}^2=\dfrac{\EWvev^2}{2M_i^2}y_i^2\,,\hspace{0.75cm}i=1,2\,.
\end{equation}

In this minimal case, $m_\nu$ is a rank-2 matrix and therefore one neutrino remains massless.
This scenario is similar to the minimal LSS, since both have only two independent Yukawas to build the Weinberg operator. 
The difference lies in the heavy sector, which has now two pairs, and the additional degrees of freedom in the $\mu$ matrix, which generates HNL mass splittings already at the leading order. 
Indeed, in contrast even to the next-to-minimal LSS, it is enough to consider the simplest case of real parameters for both $Y$ and $\mu$ to find scenarios with very suppressed or very enhanced HNL mass splittings. 
In this context, the two non-vanishing light neutrino masses can be computed analytically in terms of four quantities: the $3$ mass scales $\mutilde_i$ and the scalar product $\sigma\equiv \mathbf{u}_1^\dagger \mathbf{u}_2\in [0,1)$.
More precisely, the eigenvalues $\lambda_\pm$ of the $m_\nu$ matrix in Eq.~\eqref{mlightMinimalISS} are given by,
\begin{equation}
    \lambda_{\pm}=\dfrac{1}{2}\left(\mutilde_1+\mutilde_2+2\mutilde_3\sigma\pm\sqrt{\left(\mutilde_1+\mutilde_2+2\mutilde_3\sigma\right)^2-4\left(\mutilde_1\mutilde_2-\mutilde_3^2\right)\left(1-\sigma^2\right)}\right).
    \label{2ISS:eigenvalues}
\end{equation}
Notice that, while $\lambda_+$ can be always considered positive\footnote{We can always flip the sign of  $\mutilde_1+\mutilde_2+2\mutilde_3\sigma$, so that it is positive, rephasing by $e^{i\pi/2}$ the fields $N'_1$ and $N'_1$ together with $N_1$ and $N_2$.} and thus directly identified as the heaviest active neutrino mass, the sign of $\lambda_-$ will depend on the sign of the second term in the square root. This term is proportional to the determinant of the $\mu$-matrix:
\begin{equation}
    \mutilde_1\mutilde_2-\mutilde_3^2=U_1^2 U_2^2\cdot \text{det}\mu\,,
\end{equation}
so, depending on the sign of $\det \mu$, we have two distinct cases for the physical light neutrino masses $m_\pm$:
\begin{align}
m_+&=\lambda_+\,,
& 
m_-&= {\rm sign}\big({\rm det}\, \mu\big)\, \lambda_-\,,
\end{align}
which will lead to qualitatively different results in the following.

On the other hand, the HNL mass splittings are given by
\begin{equation}
    \Delta M_i = \mu_i-\mutilde_3\sigma=\mutilde_i/U_i^2-\mutilde_3\sigma\,,
    \label{2ISS:DeltaM}
\end{equation}
where the first term arises from the leading order term of Eq.~\eqref{type1SSMN} and the second from the subleading term of $\mathcal O(\Theta^2 M_M)$.
In general, it is enough to take into account just the first term, as the second one is suppressed by the small mixing angles, although we will discuss the validity of this approximation later. 

It is important to notice that light neutrino masses are only able to fix the $\mutilde_i$ quantities, and thus, unlike the LSS scenario, it will not be possible to compute the mass splittings in a mixing-independent way, {\it i.e.}, we will only be able to fix the mass splitting times the mixing squared, $\Delta M_i\cdot U_i^2$.
This was already expected from the naive estimate in Eq.~\eqref{DeltaM_ISS}, which we are generalizing now to a more realistic ISS able to accommodate oscillation data. For this reason, we scan in $\mutilde_1$ and $\mutilde_2$ and, for each point, we fix the value of $\mutilde_3$ and $\sigma$ that correctly reproduce oscillation data. 
We specifically scan both cases with positive and negative $\det \mu$, since they cannot be related by field rephasing and lead to qualitatively different results. 

We find that, for the $\det\mu>0$ case, only a relatively small region of $\left(\mutilde_1,\mutilde_2\right)$ space can be made compatible with oscillation data.
In particular, we find that $\mutilde_1$ and $\mutilde_2$ cannot be separated by many orders of magnitude, which in turn limits the values that $\Delta M_1\cdot U_i^2$ and $\Delta M_2\cdot U_i^2$ can have. Indeed, if for instance $\mutilde_2\ll\mutilde_1$, then $\mu_3$ cannot be too small in order to correctly reproduce two neutrino masses, which means $\mutilde_3\sim\mutilde_1\gg\mutilde_2$ and thus $\det\mu\propto \mutilde_1\mutilde_2-\mutilde_3^2<0$. Then, the regions in which there are strong hierarchies between $\mutilde_1$ and $\mutilde_2$, and that can thus cover the full parameter space require $\det \mu<0$ and this is the situation to which most points studied correspond to.

\begin{figure}[t!]
    \centering
    \includegraphics[width=1\textwidth]{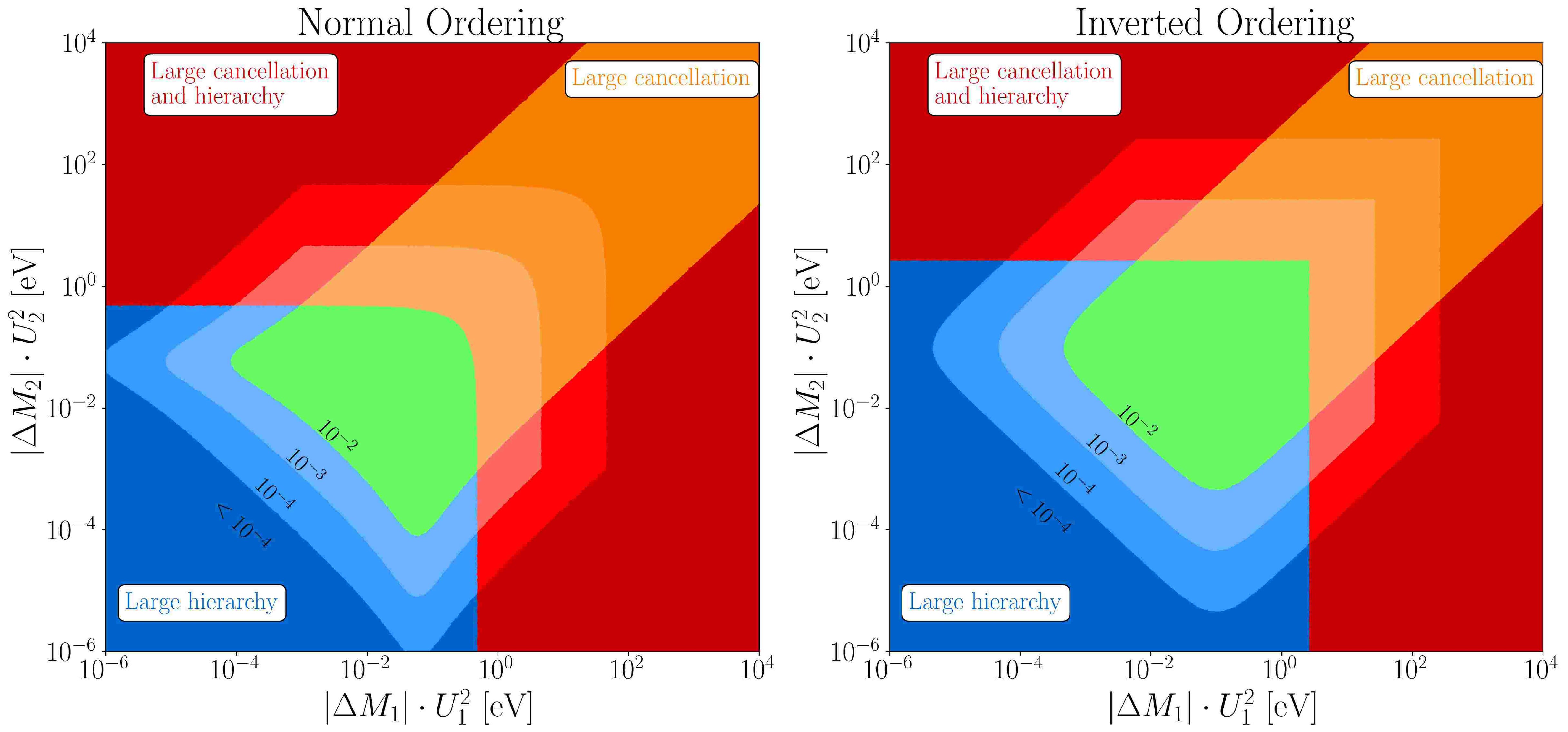}
    \caption{Different regions of parameter space of the minimal ISS compatible with oscillation data.
    In the orange area large cancellations between parameters are required and in blue area one needs a large hierarchy between parameters of the same nature. In the red area both conditions happen at the same time, while in the green area neither of them.
    In every case, darker color indicates more extreme conditions.
    Further details are given in the text.}
    \label{Fig:ISS_DegradedScatter}
\end{figure}

The results of our numerical study are shown in Fig.~\ref{Fig:ISS_DegradedScatter}, for both NO and IO.
As in the case of the LSS, we identify with different colors the various regions of the parameter space where some specific condition needs to be met.
In particular, in the orange regions the model contains large parameters that need to cancel each other in order for the light neutrino masses to remain small. In the blue area there is a large hierarchy among parameters that could be expected to be of the same order, while in the red area these two conditions happen at the same time.
Conversely, the green region does not show any specific pattern and the values for the mass splittings are not far from the naive expectations of $\mathcal O(m_\nu/U^2)$.

We will discuss the orange and blue regions in more detail, since the red one is just the intersection of these two.
The orange region corresponds to large values of $\left(\mutilde_1,\mutilde_2\right)$, so that each of the HNL pairs induces too heavy contributions to the mostly-active neutrino masses, and therefore the two contributions need to cancel each other to a large degree.
In other words, when $\mutilde_{1,2,3}$ become much larger than the physical neutrino masses, a precise cancellation to several orders of magnitude must occur between their different contributions in order to generate the correct neutrino masses. 
In order to quantify the amount of cancellation, we can compare the scale of neutrino masses to the expected mass if the three $\mutilde_{1,2,3}$ scales were not allowed to cancel each other. 
For example, from Eq.~\eqref{2ISS:eigenvalues} we can define
\begin{equation}
    c_\mu\equiv \dfrac{m_{-}}{\abs{\mutilde_1}+\abs{\mutilde_2}+2\abs{\mutilde_3\sigma}}\,,
    \label{2ISS:cmu}
\end{equation}
which will be smaller when stronger cancellations take place. 
We then identify regions with large cancellations by imposing $c_\mu < 10^{-2}, 10^{-3}$ or $10^{-4}$,  displayed in Fig. \ref{Fig:ISS_DegradedScatter} from lighter to darker orange. 
Therefore, it becomes clear that enhancing the HNL mass splittings in the minimal ISS requires of large cancellations between a priori independent parameters, in a very similar fashion to the LSS scenario in Figs.~\ref{2LSS:MultiDegradedScatters} and \ref{FIG:LSS_DegradedScatterComplex}.

On the other hand, the blue region corresponds to small $\left(\mutilde_1,\mutilde_2\right)$ values, which can be actually connected to cases with large hierarchies among the entries of the $\mu$ matrix.
Indeed, notice that in the case of small $\mutilde_{1,2}$ their contribution to neutrino masses are negligible, and it is  then $\mutilde_3$ who must be responsible of explaining their measured values. As a consequence, the smaller we make $\mutilde_1$ and $\mutilde_2$, the more off-diagonal our $\mu$ matrix becomes, which, in absence of a specific symmetry argument providing a rationale for that hierarchy, can also be considered a source of fine-tuning.
A good measure of this hierarchy is the following quantity, which has the advantage of being independent of the HNL mixings:
\begin{equation}
    h_\mu\equiv \dfrac{\mutilde_1\mutilde_2}{\mutilde_3^2}=\dfrac{\mu_1\mu_2}{\mu_3^2}\,.
    \label{2ISS:hmu}
\end{equation}
With this definition we identify areas with very hierarchical $\mu$ matrices in Fig.~\ref{Fig:ISS_DegradedScatter} through contours corresponding to $h_\mu < 10^{-2}, 10^{-3}$ or $10^{-4}$ (from lighter to darker blue).
The same conditions apply to the red contours, which satisfy the $h_\mu$ and $c_\mu$ criteria at the same time. 

Given the trend shown in Fig.~\ref{Fig:ISS_DegradedScatter}, the limit $\mutilde_{1,2}\rightarrow0$ in order to obtain an extremely small value for $\Delta M_{1,2}$, as required for instance in the Shi-Fuller mechanism, could be interesting. From the blue area, this seems possible as long as we allow for very off-diagonal $\mu$ matrices.
Furthermore, it is easy to justify such a hierarchy if the two HNLs involved in the $\mu$ matrix have opposite charge under some new symmetry, much in the same way that lepton number is invoked to have a mostly off-diagonal Majorana matrix in the low-scale seesaw scenarios under study.
Nevertheless, this argument has an important caveat. 
In Fig.~\ref{Fig:ISS_DegradedScatter}, the HNL mass splittings are shown considering only the leading order contribution, {\it i.e.}~$\Delta M_i U_i^2\approx \mutilde_i$. However, for very small $\mutilde_{1,2}$, the next to leading order term in Eq.~\eqref{2ISS:DeltaM} will dominate and generate a non-zero $\Delta M$. The specific point where the approximation breaks down depends on the values of $U_{1,2}$, so it cannot be depicted explicitly in Fig.~\ref{Fig:ISS_DegradedScatter} but has been rather included in Fig~\ref{BoundaryLNV_2HNL} in the next section.
In the $\mutilde_{1,2}\rightarrow0$ limit, $m_\nu$ in Eq.~\eqref{mlightMinimalISS} has the same form as for the minimal LSS, with light neutrino masses given by
\begin{equation}\label{offdiagISS}
    m_\pm\approx\mutilde_3\left(1\pm \sigma\right)\,,
\end{equation}
and $\Delta M_i = \mutilde_3\sigma=(m_+-m_-)/2$.
Therefore, choosing a completely off-diagonal $\mu$-matrix does not allow to obtain extremely suppressed $\Delta M_i$, as the subleading contribution prevents it from going lower than the scale of light neutrino mass splittings.  
In order to obtain smaller values, one would need a very precise cancellation between the leading and subleading contributions, which would be analogous to the cancellations between the ISS- and LSS-like contributions discussed in section \ref{Sec:2HNL} and more difficult to justify. Nevertheless, these cancellations are, a priori, possible and thus we see that already in the minimal ISS with real Yukawas and Majorana mass all values of $\Delta M_1$ and $\Delta M_2$ are allowed. We thus do not consider its generalization introducing either complex phases or additional singlets.

\section{Discussion and Conclusions}
\label{Sec:Discussion}

Low-scale seesaw scenarios -- such as the inverse or the linear seesaw -- in which neutrino masses are protected by an approximate $L$ symmetry, allow for sizable mixing of the HNLs that may lead to several interesting and potentially testable phenomenological signals. Furthermore, they may provide a connection between the origin of neutrino masses and other fundamental problems of the SM such as the source of the baryon asymmetry of the universe or the nature of its mysterious DM component. Upon integrating out the HNLs, that arrange themselves in very degenerate pseudo-Dirac pairs given the approximate $L$ symmetry, both the inverse and the linear variants lead to the same effective operators and, hence, to the same low-energy phenomenology. Nevertheless, the actual mass splitting of the pseudo-Dirac pairs is significantly different in the two scenarios. Interestingly, this splitting plays a crucial role in several phenomenological observables, from inducing $L$-violating signals at laboratory experiments to controlling the final baryon asymmetry obtained through leptogenesis. 
Being a $L$-breaking quantity, its origin and size is connected to the masses of the mostly active neutrinos. 
The precise relation, however, is different in each low-scale seesaw realization. 
This is manifest in the simplest model with only one pair of HNLs, where both quantities are related by Eqs.~\eqref{DeltaM_LSS}-\eqref{DeltaM_loopISS}.
These equations fix the size of $\Delta M$, avoiding very large or very small mass splittings. 

In this work, we have explored how these relations between the HNL mass splittings and light neutrino masses generalise to non-minimal low-scale seesaws.
We have seen that adding more HNLs provides more freedom, allowing for deviations from the naive expectations where the (tree-level) mass splittings can be arbitrarily suppressed or enhanced.
Nevertheless, these deviations come at the price of pushing the model to corners of the parameter space with large cancellations between parameters or with very particular configurations, as we have characterised in Figs.~\ref{2LSS:MultiDegradedScatters}-\ref{Fig:ISS_DegradedScatter}.

\begin{figure}[t!]
\begin{center}
    \includegraphics[width=\textwidth]{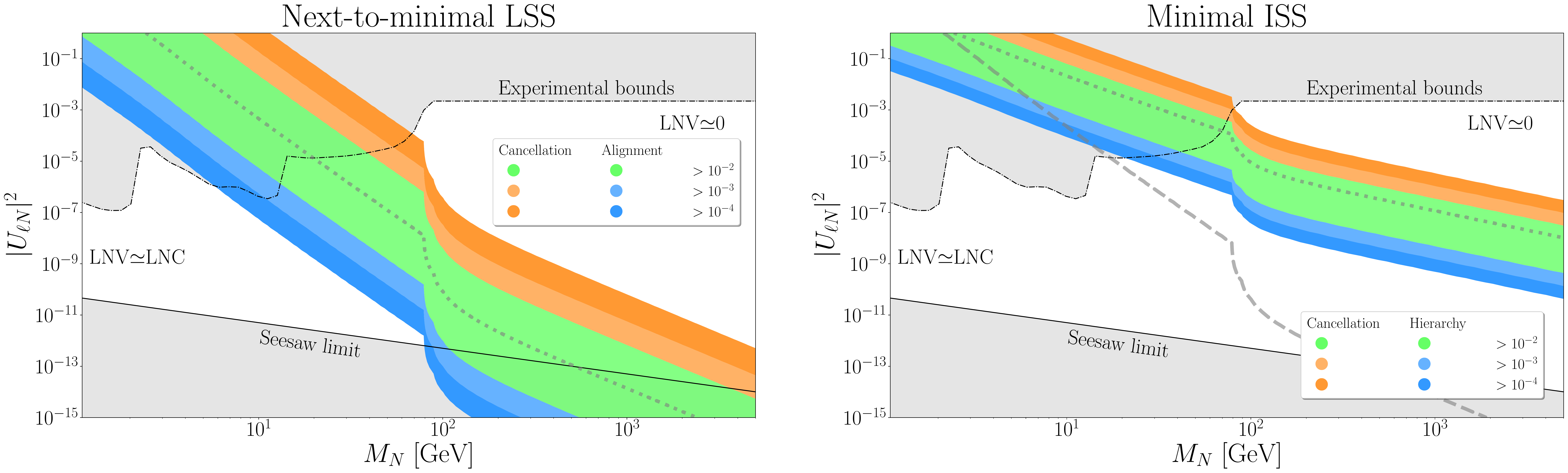}
    \caption{$\Gamma_N=\Delta M$ bands for the next-to-minimal LSS with $\mlightest=0.03$ eV (left) and for the minimal ISS (right), assuming NO in both cases. The conditions under which the width of the bands are altered are shown with the same color-coding as Figs.~\ref{2LSS:MultiDegradedScatters}-\ref{Fig:ISS_DegradedScatter}. As a comparison, we also show as gray dotted lines the lines corresponding to the minimal LSS (left) and the naive estimate for the ISS with only one HNL pair (left). For the minimal ISS plot, we have included a dashed gray line corresponding to the $\Gamma_N=\Delta M$ contour in the case of a purely off-diagonal $\mu$-matrix (that is, the dominant contribution to HNL mass splittings is the NLO of the seesaw expansion).}
    \label{BoundaryLNV_2HNL}
\end{center}
\end{figure}

In order to investigate the phenomenological implications of these deviations, we consider again Fig.~\ref{FeelingPlot} and see how it is modified upon the addition of a second pair of HNLs.
The results are shown in Fig.~\ref{BoundaryLNV_2HNL} for the case of the next-to-minimal LSS (left panel) and minimal ISS (right panel), both with 2 pairs of HNLs. 

In the simplest case with only one pair, as in Fig.~\ref{FeelingPlot}, the contour lines for $\Gamma_N=\Delta M$ delimiting the boundary between observable LNV signals and their absence, had a very sharp prediction.
With the simple extension to 2 pairs, we have shown in the previous sections that the boundary can be displaced arbitrarily up or down if sufficiently ``extreme'' configurations, either based on symmetry arguments or fine-tunnings, are allowed. Even without entering these corners of the parameter space, the extra freedom implies that the position of this boundary is not that clearly determined.
We display this variation as a band in Fig.~\ref{BoundaryLNV_2HNL}, where we have used the same color code as in Figs.~\ref{2LSS:MultiDegradedScatters}-\ref{Fig:ISS_DegradedScatter}.

In particular, we find that larger values of $\Delta M$, so that LNV signals would become observable for larger masses and mixings, requires of large cancellations between the $L$ breaking parameters of the model in both scenarios. These areas of parameter space are colored in different shades of orange depending on the level of cancellation required. On the other hand, pushing the line downwards, so that LNV signals are more suppressed, is possible for very specific configurations with (almost) perfectly orthogonal Yukawa vectors in the LSS, or with very hierarchical $\mu$ matrix in the ISS. Naively this situation might be more easily justified through symmetry arguments and, interestingly, these smaller values of $\Delta M$ may imply significant enhancements in the generation of an $L$ assymmetry in low-scale leptogenesis scenarios or even realizing the Shi-Fuller mechanism for sterile neutrino DM generation.

We again stress that all regions of the parameter space in Fig.~\ref{BoundaryLNV_2HNL} can be reached with sufficiently extreme configurations, even though we only depicted levels of cancellation or hierarchies up to $10^{-4}$. 
Notice that, due to the strong dependence of the HNL width on its mass\footnote{Scaling as $U^2 m_N^5$ below the $W$ mass and as $U^2m_N^3$ above it.}, these cancellations or hierarchies must happen at a large degree in order to show sizable deviations from the naive expectations in Fig.~\ref{BoundaryLNV_2HNL}.
But the white areas of the plot are still accessible by increasing those levels accordingly. Nevertheless, these particular solutions are not stable under next-to-leading order contributions unless a new symmetry is introduced to justify and protect them.
An example of this instability is provided by the minimal ISS with purely off-diagonal $\mu$ matrix.
At leading order, the mass splittings vanish and thus the (blue) band in the right panel of Fig.~\ref{BoundaryLNV_2HNL} would reach all the way down.
Nevertheless, as discussed in Eq.~\eqref{offdiagISS}, the NLO term becomes important here, and sets the actual position of the band to the dashed line in Fig.~\ref{BoundaryLNV_2HNL}.
Going below this line would be possible but require a cancellation between the LO and NLO terms.

Another interesting conclusion from Fig.~\ref{BoundaryLNV_2HNL} is that the boundary for LNV observability is in general rather different in both panels, providing a way to distinguish experimentally these two low-scale seesaw realizations. Although all the parameter space is technically reachable in both scenarios for sufficiently extreme configurations, a region exists in between the bands for the LSS and ISS where the former predicts suppressed LNV signals while the latter does not. Indeed, the green bands of both LSS and ISS show little to no overlap in the non-excluded parameter space. Thus,  if a HNL lying within the intermediate region was discovered, it would point out to a specific low-scale seesaw realization, providing non-trivial information on the underlying high-energy theory. 

To summarise, we have explored in depth the connection between the observable light neutrino masses and the mass splitting $\Delta M$ of the pseudo-Dirac HNL pairs beyond the minimal realizations of both the linear and inverse seesaw models. We found that values of $\Delta M$ arbitrarily larger or smaller than the naive expectation from the minimal scenario are possible. Nevertheless, significant cancellations for large $\Delta M$ or very particular configurations for small $\Delta M$ are necessary. We have quantified the level of these cancellations, hierarchies and alignments, and characterised the different areas of the parameter space showing what is required for each model to reach a certain $\Delta M$. This may provide a rationale to experimentally distinguish both scenarios. Interestingly, we find that smaller values of $\Delta M$ seem possibly related to specific symmetric configurations, particularly in the LSS which already predicts smaller values than the ISS. This suggests a way to achieve the very degenerate values of $\Delta M$ required for sufficient generation of an $L$ asymmetry so as to realise the Shi-Fuller mechanism of sterile neutrino DM production. Furthermore, such a large  $L$ asymmetry would also help to reconcile the estimations of the primordial abundances with their BBN predictions. Thus, our results may serve to throw light into the high-energy completions of the model, both by providing a guide to discriminate regions of the LSS and ISS parameter spaces through observations and by identifying some potential underlying fundamental symmetries.

\medskip

\paragraph{Acknowledgments.}
 The authors thank Jean-Loup Tastet, Manuel Gonz\'alez-L\'opez and Josu Hern\'andez-Garc\'ia for fruitful discussions during the initial phase of this project, and Jacobo L\'opez Pav\'on and Luca Merlo for very illuminating discussions.  
This project has received support
from the European Union’s Horizon 2020 research and innovation programme under the
Marie Skłodowska-Curie grant agreements No 860881-HIDDeN and No 101086085 - ASYMMETRY, and from the Spanish Research Agency (Agencia Estatal de Investigaci\'on) through the Grant IFT Centro de Excelencia Severo Ochoa No CEX2020-001007-S and Grant PID2019-108892RB-I00 funded by MCIN/AEI/10.13039/501100011033.
XM acknowledges funding from the European Union’s Horizon Europe Programme under the Marie Skłodowska-Curie grant agreement no.~101066105-PheNUmenal. The work of DNT was supported by the Spanish MIU through the National Program FPU (grant number FPU20/05333).

\appendix

\section{Scanning the next-to-minimal LSS with real Yukawas}
\label{Appendix:LSSRealScan}
In this appendix we sketch the procedure through which we have scanned the parameter space of the 2-pair-LSS in absence of complex phases.
Our goal here is to present a systematic but efficient way of scanning the parameter space that is compatible with light neutrino masses and mixings, which are generated via the mass matrix in Eq.~\eqref{LSSmnu}.

The LSS with two pairs has a rather large number of dimensions in its parameter space: even in the case in which the Yukawas are real, the physical basis contains $12$ independent couplings plus $2$ mass scales for each of the two pairs. However, for the light neutrino mass matrix only $10$ combinations of the parameters are relevant, {\it i.e.}~the $4$ unitary Yukawa directions ($\mathbf{u}_1, \mathbf{v}_1, \mathbf{u}_2, \mathbf{v}_2$), which are parametrised by $2$ angles each, and the $2$ light mass scales ($\epsilon_1, \epsilon_2$). Moreover, the correct reproduction of neutrino masses (assuming some hypothesis for $\mlightest$) and the PMNS matrix (without the CP phase) set $6$ independent constraints. Therefore, we are left with only $4$ parameters in which we randomly scan.

We will scan in different combinations of parameters, depending on which region of mass-splitting space we are, by means of solving the upper triangular part of the matrix equation:
\begin{equation}
    \UPMNS^T m_\nu \UPMNS=m_\nu^{\text{diag}}\,,
    \label{RealScanEquation}
\end{equation}
where $m_\nu$ is constructed from the model parameters as in Eq.~\eqref{LSSmnu}, $\mdiag_\nu=\text{diag}\left(m_1,m_2,m_3\right)$ and $\UPMNS$ is the usual parametrisation of the PMNS matrix:
\begin{equation}
    \UPMNS=
    \begin{pmatrix}
        c_{12}c_{13}&s_{12}c_{13}&s_{13}e^{-i\deltaCP}\\
        -s_{12}c_{23}-c_{12}s_{23}s_{13}e^{i\deltaCP}&c_{12}c_{23}-s_{12}s_{23}s_{13}e^{i\deltaCP}&s_{23}c_{13}\\
        s_{12}s_{23}-c_{12}c_{23}s_{13}e^{i\deltaCP}&-c_{12}s_{23}-s_{12}c_{23}s_{13}e^{i\deltaCP}&c_{23}c_{13}
    \end{pmatrix},
\end{equation}
with $\deltaCP=0$ for this real case. Notice, however, that $m_\nu$ is a Majorana mass matrix, thus the eigenvalues that appear in the diagonal entries upon diagonalisation are not necessarily positive. In general, one needs to add a diagonal Majorana phase matrix in order to keep them positive, which in this real case contains just signs. 
Nevertheless, these signs are not relevant for our purposes, and it is just enough to solve Eq.~\eqref{RealScanEquation} by considering the absolute value of the diagonal entries.

In order to explore the different regions in $\Delta M$ space, we have used different priors. For the $\Delta M\ll m_\nu$ we have used a log-uniform prior in the scalar products $\rho_{1,2}$; for the intermediate region $\Delta M\sim m_\nu$, we have used a uniform prior in the angles that parametrise the unitary Yukawa directions; and finally, for the $\Delta M\gg m_\nu$ we have used a log-uniform prior in the two mass scales $\epsilon_{1,2}$.

\section{Parametrisation of the complex next-to-minimal LSS}
\label{Appendix:AvelinoParametrisation}
For the purpose of scanning the parameter space of the complex Yukawa case, we will follow the parametrisation for a general neutrino mass matrix presented in~\cite{Cordero-Carrion:2019qtu} and particularise it to the next-to-minimal LSS. This parametrisation is specially convenient for the linear seesaw, since this kind of model features two distinct Yukawa coupling matrices. These are expressed as:
\begin{align}
		Y=&\dfrac{1}{i\EWvev}U' \sqrt{\mdiag} A^T W^T \sqrt{M}\,,\\
		Y'=&\dfrac{1}{i\EWvev}U'\sqrt{\mdiag}B^T W^\dagger\sqrt{M}\,,
\label{Appendix:parametrisation}
\end{align}
with the following parametrization matrices:

\begin{itemize}
	\item $W$ is a $2\times2$ unitary matrix such that:
	\begin{equation}
		W=
		\begin{pmatrix}
			e^{i\varphi_1}&0\\0&e^{i\varphi_2}
		\end{pmatrix}
		\begin{pmatrix}
			\cos\alpha&\sin\alpha\\-\sin\alpha&\cos\alpha
		\end{pmatrix}
	\begin{pmatrix}
		1&0\\0&e^{i\phi}
	\end{pmatrix}.
	\end{equation}
	In fact, the phases $e^{i\varphi_1}$ and $e^{i\varphi_2}$ are not physical, since they just rephase $\mathbf{u}_i\rightarrow e^{i\varphi_i}\mathbf{u}_i$ and $\mathbf{v}_i\rightarrow e^{-i\varphi_i}\mathbf{v}_i$, which is precisely the transformation we get if we rephase $N_i$ and $N'_i$ by $e^{i\varphi_i}$. Thus these phases encode this field-redefinition freedom and, thus, can always be chosen such that $\rho_i=\mathbf{u}_i^\dagger \mathbf{v}^{\phantom\dagger}_i\in \mathbb{R}$.
	
	\item The $A$ matrix is a product of an upper-triangular $2\times2$ matrix $T$, complex but with real diagonal entries, and an additional matrix $C_1$:
	\begin{equation}
		A=TC_1\,.
	\end{equation}
In our $2$-pair scenario, these matrices have the following form:
\begin{align}
	T&=\begin{pmatrix}
		a&c\\0&b
	\end{pmatrix}\hspace{1.30cm}\text{with}\hspace{0.5cm}a,b\in \mathbb{R},\hspace{0.25cm}c\in\mathbb{C}\,,
\\[1ex]
C_1&=\begin{pmatrix}
	z_1&1&0\\z_2&0&1
\end{pmatrix}\hspace{0.65cm}\text{with}\hspace{0.5cm}1+z_1^2+z_2^2=0,\hspace{0.25cm}z_1,z_2\in\mathbb{C}\,.
\end{align}

A possible way of parametrising the $C_1$ matrix is:
\begin{equation}
	z_1=i\cos z\,,\hspace{0.75cm}z_2=i\sin z\,,\hspace{1cm}\text{with}\hspace{0.35cm}z\in\mathbb{C}\,.
\end{equation}
\item The $B$ matrix is built from the previous matrices ($T$ and $C_1$), a complex antisymmetric $2\times2$ matrix ($K$), and the $C_2$ matrix.
\begin{equation}
	B=\left(T^T\right)^{-1}\left[C_1C_2+KC_1\right].
\end{equation}

Again in our case, $K$ and $C_2$ take the following form:
\begin{align}
		K&=\begin{pmatrix}
			0&d\\-d&0
	\end{pmatrix}\hspace{0.75cm}\text{with}\hspace{0.5cm}d\in\mathbb{C}\,,
\\
C_2&=\begin{pmatrix}
	-1&0&0\\0&1&0\\0&0&1
\end{pmatrix}.
\end{align}
\end{itemize}


\bibliographystyle{JHEP} 
\bibliography{biblio}

\end{document}